\documentclass[journal]{IEEEtran}
\ifCLASSINFOpdf
\else
\fi
\usepackage{array}
\usepackage{amsmath}
\usepackage{mdwmath}
\usepackage{mdwtab}
\usepackage{multirow}
\usepackage{amsfonts}
\usepackage{amssymb}
\usepackage{mathtools}
\usepackage{amsbsy}
\usepackage{graphics}
\usepackage{algorithm}
\usepackage{algpseudocode}
\usepackage[caption=false,font=footnotesize]{subfig}

\begin{document}
\title{Centralized and Distributed Intrusion Detection for Resource Constrained Wireless SDN Networks}

\author{Gustavo~A.~Nunez~Segura,~\IEEEmembership{Student~ Member,~IEEE,}
        Arsenia~Chorti,~\IEEEmembership{Senior Member,~IEEE,}
        and~Cintia~Borges~Margi,~\IEEEmembership{Member,~IEEE}% <-this % stops a space
\thanks{Gustavo A. Nunez Segura and Cintia Borges Margi are with Departamento de Engenharia de Computação e Sistemas Digitais, Universidade de São Paulo, São Paulo 05508-010, Brazil.}% <-this % stops a space
\thanks{Arsenia Chorti is with ETIS UMR8051, CY Université, ENSEA, CNRS, F-95000, Cergy, France.}% <-this % stops a space
\thanks{Manuscript received -- 19, 20--; revised -- 26, 20--.}}


\markboth{IEEE Internet of things Journal,~Vol.~XX, No.~XX, March~20XX}%
{Shell \MakeLowercase{\textit{et al.}}: Bare Demo of IEEEtran.cls for IEEE Journals}



\maketitle

\begin{abstract}
Software-defined networking (SDN) was devised to simplify network management and automate infrastructure sharing in wired networks. These benefits motivated the application of SDN in wireless sensor networks to leverage solutions for complex applications. However, some of the core SDN traits turn the networks prone to denial of service attacks (DoS). There are proposals in the literature to detect DoS in wireless SDN networks, however, not without shortcomings: there is little focus on resource constraints, high detection rates have been reported only for small networks, and the detection is disengaged from the identification of the type of the attack or the attacker. Our work targets these shortcomings by introducing a lightweight, online change point detector to monitor performance metrics that are impacted when the network is under attack. A key novelty is that the proposed detector is able to operate in either centralized or distributed mode. The centralized detector has very high detection rates and can further distinguish the type of the attack (from a list of known attacks). On the other hand, the distributed detector provides information that allows to identify the nodes launching the attack. Our proposal is tested over IEEE 802.15.4 networks. The results show detection rates exceeding $96\%$ in networks of 36 and 100 nodes and identification of the type of the attack with a probability exceeding $0.89$ when using the centralized approach. Additionally, for some types of attack it was possible to pinpoint the attackers with an identification probability over $0.93$ when using distributed detectors.
\end{abstract}


% \fi
\IEEEpeerreviewmaketitle

\section{Introduction}
\IEEEPARstart{W}{ireless} sensor networks (WSN) and Internet of things (IoT) consist of wireless sensor equipped devices that collect and relay information from physical and environmental phenomena. IoT WSN networks are known as resource constrained networks because, typically, sensor devices have processing, memory and energy limitations. Complex applications, with hundreds or thousands WSN nodes, may require a complex infrastructure, which is a challenge in constrained networks.

Software-defined networking (SDN) was devised to simplify network management and automate infrastructure sharing in wired networks \cite{Ieee2015}. These benefits motivated the application of SDN in WSN and IoT to leverage solutions for complex applications. The fusion of SDN -- WSN and SDN -- IoT are referred to as software-defined wireless sensor networks (SDWSN) and software-defined Internet of things (SDIoT), respectively \cite{Kobo2017}, \cite{8017556}. 

Network control centralization and data and control planes' separation are fundamental enablers of SDN programmability and network reconfiguration. On the other hand, these traits turn the network prone to denial of service (DoS) attacks, a vulnerability that is inadvertently passed on to SDWSNs and SDIoT \cite{8104765} \cite{8377989}. There are proposals in the literature to detect and mitigate DoS attacks in SDNs, and in fact, some of them focused on SDWSNs and SDIoT. However, related solutions are not adapted to very restricted networks, such as out-of-band connection for control packets between switches and controllers. Additionally, most works reported high detection rate only for small networks. Other shortcomings we noticed in the literature are a lack of solutions capable to detect multiple types of DoS attacks, identify the type of the attack and the attacker itself.

With these challenges in mind, we propose a novel DoS detector for constrained SDN networks based on change point (CP) detection theory. Our main hypothesis is that detecting a change in the monitored network metrics can be used as an alert for an anomaly, i.e., for intrusion detection purposes. A key novelty is that the proposed detector is able to operate in either centralized or distributed detection, which is not common in SDN-based networks. In the centralized detection, a security application monitors the control packets overhead and the data packets delivery rate of the network. If the application detects a change on the statistical properties of one of these metrics, the network is considered under attack. In the distributed detection, every node is in charge of detecting a change on its own local metrics and to inform the security application in case of a change.  Notably, the centralized detector that runs on the controller allows to \text{identify with a very high rate the attack} and further can \textit{distinguish the type of the attack} from a list of known attacks. The distributed detector that runs on individual nodes is also able to \text{detect the DoS attacks with a high rate} and further provides information that allows to \textit{identify the nodes launching the attack}.

\begin{table*}[t]
\caption{Related Work}
\label{tab:related}
\centering
\begin{tabular}{|l|c|c|c|c|c|}
\hline
Author                                  & High detection rate & \begin{tabular}[c]{@{}c@{}}Multiple types \\ of attack\end{tabular} & \begin{tabular}[c]{@{}c@{}}Attack type \\ identification\end{tabular} & \begin{tabular}[c]{@{}c@{}}Resource constrained \\ networks\end{tabular} & \begin{tabular}[c]{@{}c@{}}Attacker \\ identification\end{tabular} \\ \hline
Bhunia and Gurusamy \cite{8215418}      & \checkmark                   &                                                                     &                                                                       &                                                                 &                                                                    \\ \hline
Jia \textit{et al.} \cite{9090824}      & \checkmark                   & \multicolumn{1}{l|}{}                                               & \checkmark                                                                     & \multicolumn{1}{l|}{}                                           & \multicolumn{1}{l|}{}                                              \\ \hline
Ravi and Shalinie \cite{8993716}        & \checkmark                   &                                                                     &                                                                       &                                                                 & \checkmark                                                                  \\ \hline
Yin \textit{et al.} \cite{8352645}      &                     &                                                                     &                                                                       & \checkmark                                                               & \checkmark                                                                  \\ \hline
Miranda \textit{et al.} \cite{8998393}  &                     & \checkmark                                                                   &                                                                       & \checkmark                                                               &                                                                    \\ \hline
Wang \textit{et al.} \cite{WANG2018119} &                     & \checkmark                                                                   & \checkmark                                                                     & \checkmark                                                               & \checkmark                                                                  \\ \hline
Our proposal                            & \checkmark                   & \checkmark                                                                   & \checkmark                                                                     & \checkmark                                                               & \checkmark                                                                  \\ \hline
\end{tabular}
\end{table*}

We measured the performance of both approaches on the IT-SDN framework \cite{8805072}, simulating new-flow and neighbor information types of attacks in topologies of 36 and 100 nodes, when all the sensor nodes were emulated as Tmote sky. Our contributions are listed below:
\begin{enumerate}
    \item We developed DoS detectors suitable for restricted networks (IEEE 802.15.4).
    \item Our detectors do not need training data (as for example do machine learning based detectors), they require only $200$ samples of the monitored time series when the network is not under attack to extract its statistics.  
    \item We studied the parameterization of the centralized detector to optimize the detection speed versus the detection rate and studied the trade-off between the two. The quickest detector achieved an attack identification rate of more than $89\%$, increased to $99\%$ for less agile detectors. 
    \item The decentralized detector is so lightweight that can run on each individual Tmote sky node in the network, which allowed us to identify the region in which the attack is launched, or even, the attacker itself with a probability exceeding $93\%$.
\end{enumerate}

The remaining of the paper is organized as follows. In Section \ref{sec:relatedwork} the state of the art is summarized while in Section \ref{sec:problem} the intrusion scenario is explained. Section \ref{sec:itsdn} overviews SDWSN security vulnerabilities, while Section \ref{sec:prediction} presents the mathematical background for the change point detector. In Section \ref{sec:centralized-approach} and \ref{sec:distributed-approach} we present the centralized and distributed detectors, respectively, and discuss their results. Section \ref{sec:tracking} presents the overall attacker detection strategy and the  discussion of the performance. Lastly, Section \ref{sec:conclusion} concludes the paper.

\section{Related Work} \label{sec:relatedwork}

In this section, we analyze works that propose solutions for resource constrained SDN-based networks. Our focus is on proposals targeting DoS attacks detection and identification. The analysis is based on DoS attacks detection and identification accuracy, type of DoS attacks detected, and the consideration of resource constraints.

Table \ref{tab:related} summarizes the main performance metrics of the related work and our proposal. We chose five metrics for the comparison: i) the ability to achieve high detection rates i.e., equal or greater than $90\%$; ii) multiple types of attack detection; iii) type of attack identification; iv) resources limitations; and v) attacker identification. One general comment is that most of the previous works reviewed here are OpenFlow-based, which limits their use in networks composed of constrained nodes because of limited frame sizes, memory constraints, and lack of dedicated control channel. Because of this, some papers did not include real devices emulation or testbeds.

Bhunia and Gurusamy \cite{8215418}, Ravi and Shalinie \cite{8993716}, and Jia \textit{et al.} \cite{9090824} proposals have in common that all of them used machine learning techniques to detect DoS attacks, and also all of them obtained high detection rate results, i.e., higher than $90\%$. On the other hand, none of these three proposals considered resource constraints or were evaluated on restricted networks. The main reason is because these are OpenFlow-based or require a high traffic of packets to monitor the network. About the other metrics, Jia \textit{et al.} \cite{9090824} proposed an attack type identification algorithm and Ravi and Shalinie \cite{8993716} proposed an attacker identification mechanism.

Yin \textit{et al.} \cite{8352645}, Miranda \textit{et al.} \cite{8998393}, and Wang \textit{et al.} \cite{WANG2018119} proposals have in common that all of them considered resource constrained networks, but on the other hand did not attain high detection rates. Concerning the other metrics, Miranda \textit{et al.} \cite{8998393}, and Wang \textit{et al.} \cite{WANG2018119} proposed multiple types of attack detection, and Yin \textit{et al.} \cite{8352645} and Wang \textit{et al.} \cite{WANG2018119} proposed an attacker identification algorithm.

The main shortcoming in the state of the art is the tradeoff between detection rate and resources to execute the DoS attack detector. The proposals that attained high detection rate were not suited for resource constrained networks, and proposals that considered resource limitations did not attain high detection rates. As shown in Table \ref{tab:related}, our solution obtained high detection rates while it is well suited for resource constrained networks. Additionally, our solution was able to detect different types of DoS attack, identify the type of the attack with high probability, and identify the area where the attacker is located, or even the attacker itself. Our proposal was the only one fulfilling the five metrics.

The present study builds upon our previous works in \cite{OJIOT2019gnunez}, \cite{icc-2020}, and \cite{latincom-2020}. In the first work \cite{OJIOT2019gnunez} we analyzed the impact of different types of attacks on various performance metrics and identified the data packet delivery and the control overhead rates as the most impacted. In \cite{icc-2020}, on the other hand, we proposed a universal CP DoS detector that combined an offline and an online detector. Lastly, in \cite{latincom-2020} we moved to an entirely online multimetric CP detector, which used two centralized CP detectors independently optimized for different types of attack. This strategy allowed us to obtain high detection rates for different attacks in topologies up to 100 nodes, and, more importantly, to identify the type of the attack we are detecting. It is also important to mention that unlike machine learning based detectors we do not need large training data sets; as will be shown, around 200 samples of the metric monitored suffice to extract its statistical characteristics. This turns our solution more lightweight and general, well  suited for resource constrained networks. 

\begin{table*}[htb]
\caption{WSN Motes Specifications}
\label{tab:motes}
\centering
\begin{tabular}{|l|c|c|c|c|}
\hline
\textbf{Platform}                                         & \textbf{\begin{tabular}[c]{@{}c@{}}Microprocessor\\ model\end{tabular}} & \textbf{\begin{tabular}[c]{@{}c@{}}Clock speed\\ (MHz)\end{tabular}} & \textbf{\begin{tabular}[c]{@{}c@{}}Flash memory\\ (kB)\end{tabular}} & \textbf{\begin{tabular}[c]{@{}c@{}}RAM \\ (kB)\end{tabular}} \\ \hline
TelosB                                                    & MSP430                                                                  & 8                                                                    & 48                                                                   & 10                                                           \\ \hline
SensorTag                                                 & ARM Cortex-M3                                                           & 48                                                                   & 128                                                                  & 20                                                           \\ \hline
RE-Mote                                                   & ARM Cortex-M3                                                           & 32                                                                   & 512                                                                  & 32                                                           \\ \hline
Raspberry Pi $3$ & 4 x ARM Cortex-A53                                                      & 1200                                                                 & SD card                                                              & 1 000 000                                                         \\ \hline
\end{tabular}
\end{table*}

In the current work, we further extended our previous works and proposed a distributed DoS attack detection approach based on metrics collected and analyzed on every node, including the transmitting time, processing time, etc., hinting to the possibility of intrusion detection at PHY and it's potential incorporation with physical layer security solutions \cite{Chorti-Hollanti}. From this, we were able to implement our security solution in either centralized or distributed detection according to the network resources. The centralized approach requires more bandwith while the distributed requires more of the nodes' memory.

% Lastly, we evaluated that using the distributed detectors we are able to identify the attacker in new-flow based type of attacks. 

% With our proposal we addressed the resource constraints and high detection rate shortcomings in medium and large networks, but also we provide mechanisms to identify the type of the attack and the location of the attackers in new-flow based type of attacks.  

\section{SDWSN Vulnerabilities Overview} \label{sec:problem}
SDNs have a centralized architecture, where a controller, or multiple controllers and their interfaces constitute the control plane and are in charge of the of network's configuration \cite{McKeown2008}. Because of this centralization, the controller has a global view of the network topology and also can have access to traffic and performance information. 

In terms of security, SDNs have advantages and disadvantages. The access to network's traffic and performance data along with the controller's global view, is a combination that has been used to develop new security strategies \cite{7226783}. Based on a centralized traffic analysis and security policies, the controller has an important role to determine if the network is under attack and to reconfigure the network to mitigate the impact. On the other hand, SDNs are entirely controller-based. This means, if the controller is compromised, the control plane is compromised, therefore, the whole network is compromised as well. For this reason, the controller is tagged as a single point of failure, which turns SDN-based networks prone to DoS attacks \cite{SINGH2020509} \cite{7593247}.

In SDNs, the attackers can reach the control plane directly through the controller or through network devices. An attacker can flood the network with control packets that will be forwarded to the controller, exhausting its processing and communication resources. In the same way, an attacker can mislead other nodes in the network, inducing them to communicate with the controller at the same time, similar to a flooding attack. For example, an attacker sends several data packets tagged with an unknown flow identifier. The neighbouring nodes receiving the packet will check on their routing table to match the packet's flow identifier with a rule but without success, therefore the nodes will request a flow rule from the controller. This type of attack impacts both the controller and the network devices' resources, leading to a compromise of the entire network. 

In the case of SDWSN and IoT, the previous scenario is critical since network devices are resource constrained. To have a better idea about these constraints, we summarized some IEEE 802.15.4 compliant platforms in Table \ref{tab:motes} and compare them with Raspberry Pi $3$ specifications, a small single-board computer. Because of resources constraints, to deal with saturation attacks, resource exhaustion, and complex security mechanisms become challenging. As shown in our previous work  \cite{OJIOT2019gnunez}, a new-flow-based attack \cite{ShinSeungwon} is able to increase the number of control packets per minute between 16\% and 127\% when there is only one attacker in the network, and the impact increases when increasing the number of attackers. Additionally, after several repetitions, this attack can saturate the neighbors' flow tables. In \cite{OJIOT2019gnunez}, we also investigated the impact of one topology discovery based attack \cite{7534866} in SDWSN and showed that one attacker in the network was able to reduce the data packets delivery rate between 5\% to 18\%. Multiple attackers executing this attack were able to reduce the data packets delivery rate between 20\% to 60\%.

\subsection{IT-SDN and DoS attacks} \label{sec:itsdn}
IT-SDN \cite{8805072} is an SDWSN framework composed of the application layer, the control layer, and three communication protocols: the Southbound protocol, the Neighbor Discovery protocol, and the Controller Discovery protocol. 

The sensing layer is composed of the wireless sensor devices used to collect data from the environment and relay data to sinks, this means the wireless sensor network itself. The control plane is composed of the control servers in charge of taking and installing routing decisions in the sensing layer devices. The Southbound protocol defines the message formats for communication between the WSN and the controller. The Neighbor Discovery and the Controller Discovery protocols are often executed as a joint operation. All nodes in the network use the Controller Discovery protocol to find a route to reach the controller and use the Neighbor Discovery protocol to collect neighborhood information to then send it to the controller. 

The controller uses the neighborhood information provided to calculate routes according to a set of policies and procedures. Then, the controller installs the routing rules on every node's flow table. IT-SDN's flow table is composed of four columns: matching criteria, action taken, action parameter, and flow usage. The matching criteria is an address or a flow ID. The actions are: forward packet, drop packet, or receive packet. The action parameter is typically the next hop and the flow usage is assessed as the number of updates since the entry was installed.  

The Southbound protocol is composed of six packet types: flow request, flow setup, flow ID register, acknowledgement, neighbor report, and data packet. Next we define only the ones involved in our scope. The WSN's nodes use the flow request packet to ask the controller about an unknown route and the controller replies with a flow setup packet that contains the route configuration. Moreover, the controller can change any route configuration using this packet whenever a route calculation changes. The neighbor report packet contains the sender's neighborhood information. The controller uses this information to update the graph and recalculate routes. The nodes send a neighbor report to the controller on one of three conditions: the node detects one or more new neighbors, the node detects one or more nodes are no longer his neighbors, or there is a significant change on one or more neighbors' routing metric. A significant change is defined as a percentage of change of the current metric that can be defined according to the application. 

In this work we tested our proposal when the network is under two different attacks: new-flow-based attack and neighbor information type of attack. The new-flow-based attack \cite{ShinSeungwon} is characterized by the flooding of packets with the objective to include new flows in the networks, and this is why this attack is commonly studied in SDN. The neighbor information attack targets important information sent from all nodes to the controller to calculate routing rules. This attack has not been explored widely in the state of the art and from our previous work \cite{OJIOT2019gnunez} we observed it significantly disturbs the data and control packets delivery rate. Based on IT-SDN characteristics, we adapted these two attacks to target its security vulnerabilities, dubbed in the rest of this paper as the \textbf{false data flow forwarding (FDFF)} and the \textbf{false neighbor information (FNI)}.  

\begin{enumerate}
\item The \textbf{false data flow forwarding (FDFF)} targets the controller via network's devices. First, the attacker sends data packets with unknown flow identifiers to its neighbors. The neighbors receive the packet and check the flow table to determine the action required, without success, thus ask a rule to the controller by sending a flow rule request packet. The controller receives this packet, calculates the rule and replies sending a flow setup packet. Fig. \ref{fig:attack-1} shows the packets exchange during this attack, which aims at increasing the network's packet traffic and the controller's and neighbors' processing overhead.  

\begin{figure}[t]
    \centering
    \includegraphics[width=0.40\textwidth]{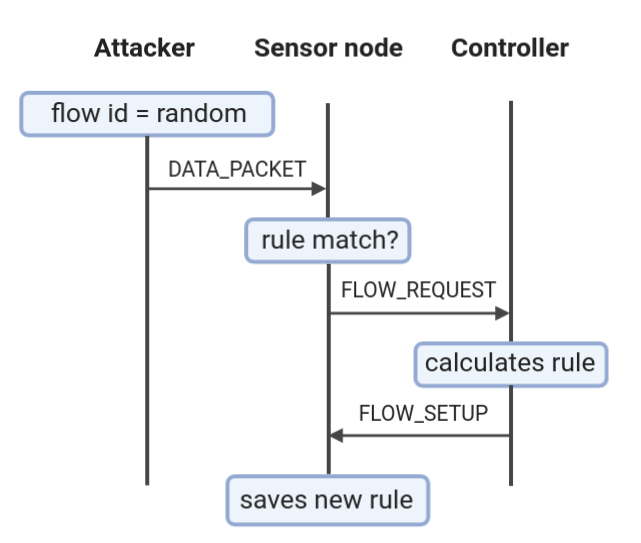}
    \caption{False data flow forwarding attack: the attackers inside the network send data packets to their neighbors using random or unknown identifiers. The sensor nodes request a rule to the controller to treat this packet, the controller calculates the rule and send it to the sensor node}
    \label{fig:attack-1}
\end{figure}

\item The \textbf{false neighbor information (FNI)} attack modifies the packets that contain neighbor information. The attackers do not intercept the neighbor information packets but modify the ones that use them to reach the controller. When receiving a neighbor information packet, the attacker modifies either the routing metric or node identification number, then the packet continues its normal route to reach the controller. The packets exchange diagram for this attack is depicted in Fig. \ref{fig:attack-3}. This attack leads the controller to mistreat false information as true and will send erroneous routing rules to the nodes. 

\begin{figure}[t]
    \centering
    \includegraphics[width=0.40\textwidth]{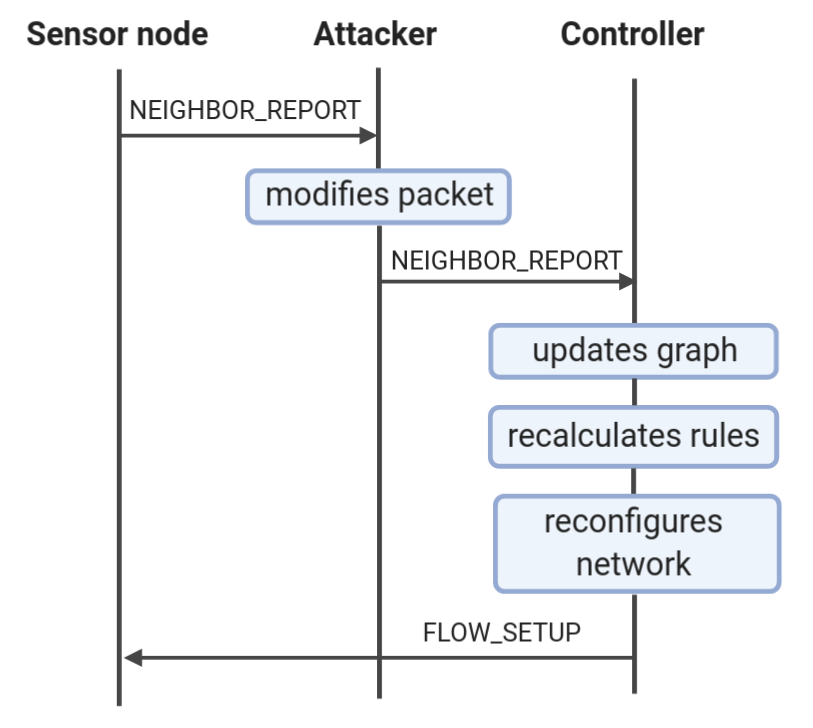}
    \caption{False neighbor information attack: the sensor node sends a neighbor report to the controller and the attacker in the route (in the case there is one) modifies the neighborhood information before forwarding the packet to the controller}
    \label{fig:attack-3}
\end{figure}
\end{enumerate}

\section{Online Change Point Detection} \label{sec:prediction}

In this section we explain the basic of change point (CP) analysis and the algorithm we used for DoS attack detection in SDWSN. Generally, change point problems have been phrased as hypothesis tests. The null hypothesis is established to represent structural stability of the process, while the alternative hypothesis contains one or multiple change points. The test statistics may be viewed as two-sample tests adjusted for the unknown break location, thus leading to max-type procedures. Often asymptotic relationships are derived to obtain critical values for the tests. After the null hypothesis is rejected, the location(s) of the break(s) need(s) to be estimated \cite{alexander-timeseries}. 

It was shown in \cite{OJIOT2019gnunez} that FDFF attacks induce substantial changes in mean control packet rates while FNI attacks induce important changes in  mean data packets delivery rates. From this analysis, we formulated the attack detection problem as a hypothesis test, examining whether a change has occurred in the mean value of the time series observed for these two metrics. 

Regarding the CP methodologies that incorporate the serial dependence of the observations into the statistical analysis, we can distinguish between parametric and non-parametric approaches. Focusing on non-parametric anomaly detection, i.e., without relying on assumptions regarding the underlying statistical model, we note it has typically been considered for the detection of anomalies in networks' traffic. As an example, Tartakovsky \textit{et al.} \cite{6380529} proposed an algorithm for anomaly detection in computer network traffic,  Wang \textit{et al.} \cite{1388279} proposed a a cumulative sum (CUSUM) based proposal for the detection of SYN attacks, and Skaperas \textit{et al.} \cite{8988163} \cite{8835019} used mean change point analysis to detect anomalies on video content popularity. 

In this work, we employed a CUSUM based algorithm to detect changes in the mean value of control overhead and data packets delivery rate time series. This decision allowed us to alleviate the need for any parametric model with respect to the impact of the attack \cite{icc-2020}. Then, in \cite{latincom-2020} we proposed two major novelties in the detector: first, we moved to a purely online detector, unlike \cite{icc-2020} in which a hybrid offline-online algorithm was presented; secondly, we monitored in parallel multiple metrics, increasing the detection vector space to different types of attack and provided a probabilistic identification of the type of the attack.

To outline the online CP algorithm, let $\{X_{n}:n \in \mathbb{N}\}$ be the time series of the metric monitored. Using Wold's theorem we can assume that, for $X_1,...,X_N$, each sample is expressed as $X_n=\mu_n+Y_n$, where $\{\mu_n, n\in \mathbb{N}\}$ is the mean of the time series and $\{Y_n:n \in \mathbb{N}\}$ is a random zero mean term, so that we can rewrite $X_{n}$ as:
\begin{equation}\label{eq:11}
X_n= \begin{cases} 
         \mu+Y_n, \hspace{11mm} n=1,\ldots,m+k^*-1 \\
\mu+Y_n+I, \hspace{5mm} n=m+k^*,\ldots 
  \end{cases}
\end{equation}
where $k^*\in\mathbb{N^*}$ represents the unknown time of change and $\mu$, $I\in\mathbb{R}^{r}$ represent the mean parameters before and after $k^*$, respectively. In the present we assume a period of no change in the mean of at least $m$ samples, i.e., during the first $m$ samples of our observation there is no change so that: 

\begin{equation}
\mu_1=\ldots=\mu_m   
\end{equation}
 
During this period, our detector ``learns'' in real-time the statistics of the observed time series, and, the mean value in particular. Finally, the statistical hypothesis test is articulated as,

\begin{equation}\label{eq:13}
\begin{aligned}
H_0&: I=0\\
H_1&: I\neq0.
\end{aligned}
\end{equation}

The on-line sequential analysis belongs to the category of stopping time stochastic processes. In general, a chosen on-line test statistic ${TS_{on}(m,l)}$ and a given threshold $F(m,l)$ define the stopping time $\tau(m)$: 
\begin{equation}\label{eq:14}
\tau{\left(m\right)}= \begin{cases} 
         \min\lbrace{l\in{\mathbb{N}}: {TS_{on}(m,l)}{\geqslant{F(m,l)}}}\rbrace,\\
\infty, \text{ if } {TS_{on}(m,l)}{<{F(m,l)}} \text{ }\forall{l\in{\mathbb{N}}},
   \end{cases}
\end{equation}
implying that $TS_{on}(m,l)$ is calculated  on-line  for  every
$l$ in  the  monitoring  period. The procedure stops if the test statistic exceeds the value of the threshold function $F(m,l)$. As soon as this happens, the null hypothesis is rejected and a CP is detected. The following properties should hold for $\tau(m)$,

\begin{equation}
\lim_{m\to\infty} Pr{\left\lbrace\tau(m)<\infty|H_0\right\rbrace=\alpha}, 
\end{equation}
% \[ \lim_{m\to\infty} P{\left\lbrace\tau(m)<\infty|H_0\right\rbrace=\alpha},\]
ensuring that the probability of false alarm is asymptotically bounded by $\alpha\in\left(0,1\right)$, and,
\begin{equation}
\lim_{m\to\infty} Pr{\left\lbrace{\tau(m)<\infty|H_1}\right\rbrace=1},
\end{equation}
% \[ \lim_{m\to\infty} P{\left\lbrace{\tau(m)<\infty|H_1}\right\rbrace=1},\]
ensuring that under $H_1$ the asymptotic power is unity.
Under these conditions, $F(m,l)$ is defined as,
\begin{equation}\label{eq:15}
F(m,l)={cv_{on,\alpha}}{g}{\left({m,l}\right)},
\end{equation}
where: (i) $cv_{on,a}$ is the critical value determined from the asymptotic behavior of the stopping time procedure under $H_{0}$ by letting $m\rightarrow\infty$, (ii) and $g(m,l)$ is the weight function defined as:

\begin{equation}\label{eq:16}
g(m,l)=\sqrt{m}\left(1+\frac{l}{m}\right)\left(\frac{l}{l+m}\right)^\gamma
\end{equation}
where the sensitivity parameter $\gamma\in\left[0,1/2\right)$.

The online algorithm uses the standard CUSUM detector \cite{FREMDT201474}, with test statistic denoted by  $TS^{ct}_{on}$. Its corresponding critical value is denoted by $cv^{ct}_{on,\alpha}$ and the stopping rule by $\tau_{ct}(m)$. The sequential CUSUM detector is denoted by $E(m,l)$,
    \begin{equation}{\label{onlC}}
    E(m,l)=\left(\overline{X}_{m+1,m+l}-\overline{X}_{1,m}\right)
    \end{equation}

The standard CUSUM test is expressed as:
\begin{equation}\label{eq:17}
TS^{ct}_{on}(m,l)={l\widehat{\Omega}^{-\frac{1}{2}}_{m}}{E(m,l)},
\end{equation}
where $\widehat{\Omega}_{m}$ is the estimated long-run covariance, defined as in (4), that captures the dependence between observations. Then, the stopping rule $\tau_{ct}(m)$, is defined as: 
\begin{equation}\label{s.t1}
\tau_{ct}(m)=\min\lbrace{l\in\mathbb{N}:\Vert{TS^{ct}_{on}(m,l)}\Vert_{1}\geq{cv^{ct}_{on,\alpha}}g(m,l)\rbrace},
\end{equation}
where the $\ell_{1}$ norm is involved to modify $TS^{ct}_{on}$ so that it can be compared to a one dimensional threshold function.   
The critical value, $cv^{ct}_{on,\alpha}$, is derived from the asymptotic behavior of the stopping rule under $H_{0}$:

\begin{align} \label{eq:18}
 & \lim_{m\to \infty} Pr {\lbrace \tau(m)<\infty\rbrace} \\
 & = \lim_{m \to\infty} Pr {\left\lbrace{\sup_{1\leqslant{l}\leqslant{\infty}}}{\frac{\Vert{TS^{ct}_{on}(m,l)}\Vert_{1}}{g(m,l)}}{>cv^{ct}_{on,\alpha}}\right\rbrace}\nonumber \\
 & =Pr\left\lbrace\sup_{t\in\left[0,1\right]}{\frac{\Vert{W(t)}\Vert_{1}}{t^\gamma}}>cv^{ct}_{on,\alpha}\right\rbrace=\alpha
\end{align} 
where $W(t)$ denotes the Brownian motion with mean $0$ and variance $t$. 
The on-line critical values were computed using Monte Carlo simulations, considering that,      
    \begin{equation}\label{cv_standard}
    \begin{split}
    cv^{ct}_{on,\alpha}=\sup_{t\in\left[0,1\right]}{\frac{W(t)}{t^\gamma}},
    \end{split}
    \end{equation}

Lastly, the estimated on-line CP, $\hat{k}^*_{on}$, is derived directly from the value of the stopping time $\tau(m)$, as,
    \begin{equation}\label{on_line cp}
    \hat{k}^{*}_{on}=m+\lbrace\tau(m)|\tau(m)<\infty\rbrace  .
    \end{equation}

% \begin{align}\label{eq:18}
% \lim_{m\to\infty} Pr {\lbrace\tau(m)<\infty\rbrace}\\
% & =\lim_{m\to\infty} Pr {\left\lbrace{\sup_{1\leqslant{l}\leqslant{\infty}}}{\frac{\Vert{TS^{ct}_{on}(m,l)}\Vert_{1}}{g(m,l)}}{>cv^{ct}_{on,\alpha}}\right\rbrace} \nonumber\\
% &  =Pr \left\lbrace\sup_{t\in\left[0,1\right]}{\frac{\Vert{W(t)}\Vert_{1}}{t^\gamma}}>cv^{ct}_{on,\alpha}\right\rbrace=\alpha.
% \end{align}

Summarizing, the overall algorithm has 3 main steps:
\begin{itemize}
    \item Step 1: define the values of the quantities $m$,  $\gamma$, the confidence level $\alpha$, and set $l$.
     \item Step 2: after collecting $m$ samples of the metric, $\Gamma(m,l)$ (\ref{eq:17}) and the weight function in (\ref{eq:16}) are calculated for every $l$ on the monitoring period to then apply (\ref{eq:18}).
     \item If a CP is detected, the online process stops. Conversely, if the period $l$ ends, a new monitoring period is defined.
\end{itemize}

\section{Centralized Detection} \label{sec:centralized-approach}

% \subsection{Explanation}
In \cite{OJIOT2019gnunez}, it has been shown that FDFF and FNI attacks have a significant impact on the data packets delivery rate and the control packets overhead. A centralized intrusion detection, first proposed in \cite{icc-2020} and \cite{latincom-2020}, can be used to determine if the network is under attack based on monitoring of these two metrics; here we propose to use parallel detectors for both and to identify the type of the attack based on which detector triggers an alert first. In detail,  an attack is classified as a FDFF or a FNI attack based on the following reasoning: 
\begin{enumerate}
    \item If a CP is detected in the mean value of the data packets delivery rate or control packets overhead, we determine that the network is under attack;
    \item If the CP is first detected in the control packets overhead, the attack is classified as FDFF; conversely, if the CP is detected first in the data packet delivery rate, the attack is classified as FNI.
\end{enumerate}
%In the possible case where a CP is detected in both metrics, we use the first alarm triggered to decide the type of the attack.

\begin{figure}[t]
    \centering
    \includegraphics[width=0.40\textwidth]{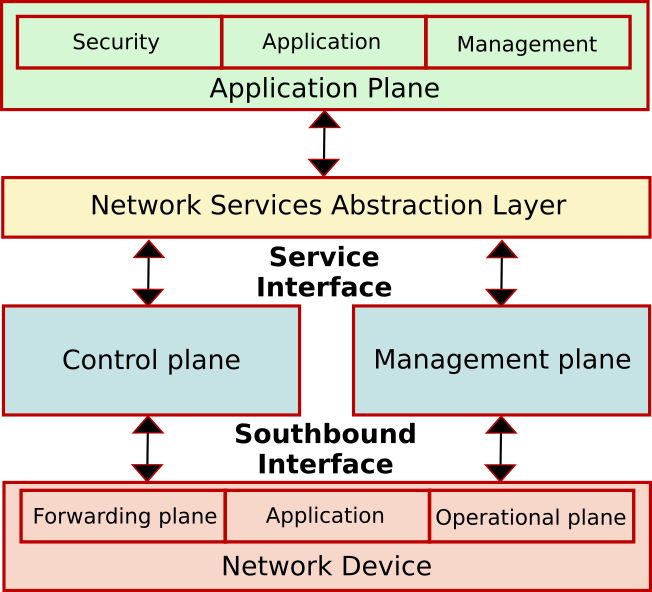}
    \caption{SDWSN architecture based on IRTF RFC 7426 document \cite{haleplidis2015software}}
    \label{fig:sdwsn_architecture}
\end{figure}

Our proposal is based on the SDN architecture proposed in the IRTF RFC 7426 \cite{haleplidis2015software}, depicted in Fig. \ref{fig:sdwsn_architecture}, for which the management plane's purpose is to ensure the network is running optimally. To accomplish this, the management plane establishes communication with the network devices using the Southbound Interface to obtain information about the network operation. Then, this information is shared with the modules in the Application Plane using the Network Services Abstraction Layer. 

We monitor the number of control packets and data packets sent by every node, and the number of data packets received by the data sink. Every node sends a packet to a management sink every two minutes, then these data are sent to the security module in the Application Plane. The security module calculates the metrics, constructs the time series and runs the CP detector algorithm explained in Section \ref{sec:prediction}. Whenever a CP is detected, the module raises an alarm indicating the metric where the CP was detected. This information could be sent to the controller to implement mitigation strategies, which is outside the scope of this work. 

\subsection{Experimental setup} \label{sec:method-centralized}

We generated a dataset comprising 480 simulations, divided in 240 simulations of FNI attacks and 240 simulations of FDFF attacks. Then, we split each subgroup in two sets: one set for parameterization to capture different trade-offs between the detection rate and the speed of detection and the other for validation. In particular, we used the first set to determine the optimal values of $\{m, \gamma\}$ (both parameters explained in Section \ref{sec:prediction}) for each type of attack and each observed metric. Then, using the values determined for $\{m, \gamma\}$, we executed the CP detector algorithm over the validation sets to evaluate the performance achieved.  We performed simulations on square grids with either 36 or 100 nodes and we varied the number of intruders (attackers) in three proportions: 5\%, 10\% and 20\% of the total of nodes in the network. 

First, we executed the algorithm on the first set for $m \in \{100, 150, 200\}$ and $\gamma \in\{0,0.15,0.25, 0.35, 0.45, 0.49\}$ to determine the values that provide the best performance for different trade-offs between the detection rate $DR$ and the detection time median $DTM$. The $DR$ is the ratio of successfully detected attacks over the total number of attacks. The $DTM$, is the median of the number of samples required to detect the attack. From that, we introduced a ``detection score'' metric to capture the relative importance that is given to the $DR$ versus the $DTM$ (which focuses on detecting changes on a signal or a time series as quickly as possible after they occur \cite{poor2008quickest}). The proposed detection score metric, $P_{DS}$, is defined as: 
\begin{equation}
    P_{DS}(A, B) = A(1-S) + B(DR), \hspace{0.5cm} A + B = 1,
\end{equation}
where $A$ and $B$ are constants to determine the relative weight of each term, and $S=\frac{DTM}{l}$ with $l$ the number of samples monitored after the attack starts. We used five combinations of $A$ and $B$, where $(A, B)\in \{(1, 0), (0.8, 0.2), (0.5, 0.5), (0.2, 0.8), (0, 1)$\}, to compare the results when prioritizing the speed of detection ($A > B$) versus when prioritizing the detection rate ($A < B$). 

During evaluation, two CP detectors were running in parallel. One detector for monitoring the control packets overhead and the other one for monitoring the data packets delivery rate. The validation set comprised both FDFF and FNI attack simulations, 50\% of each one, including all chosen topologies and attack intensity levels. In the validation stage we used the optimal pairs $(m, \gamma)$ identified for each pair $(A, B)$ to  maximize the metric $P_{DS}(A, B)$. Whenever a CP was detected, we stopped the detectors, declared the network under attack, and determined which metric triggered the detector. If the detector monitoring the control overhead was triggered first, we declared an FDFF attack, alternatively, if the detector monitoring the data packet delivery rate was triggered first, we declared an FNI type of attack. 

\begin{figure}[t]
\begin{center}
    \subfloat[FDFF attack\label{fig:a1_3_36}]{
    \includegraphics[width=0.30\textwidth]{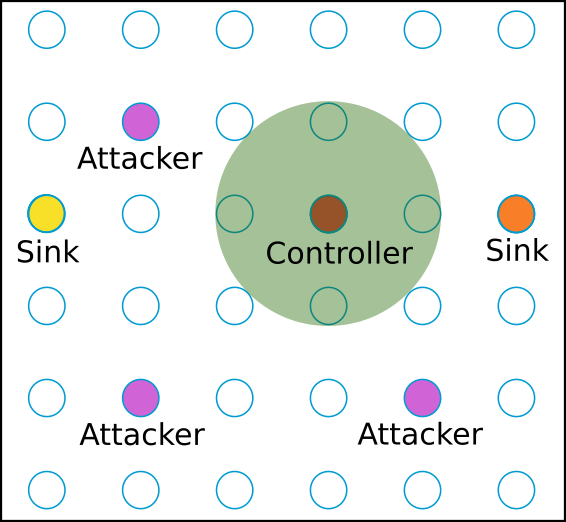}}
\hspace{\fill}
\subfloat[FNI attack\label{fig:a3_3_36}]{
    \includegraphics[width=0.30\textwidth]{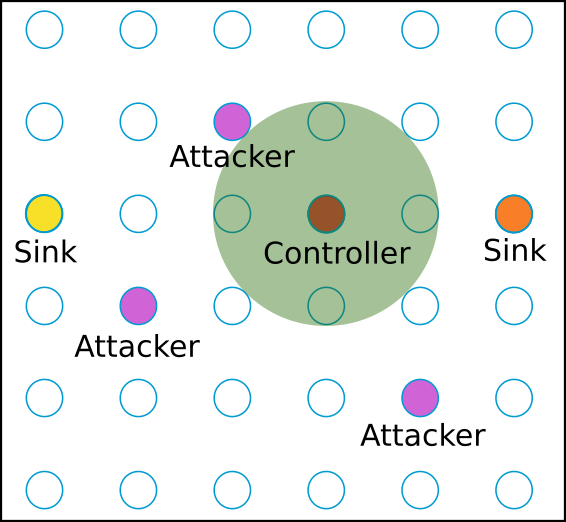}}
\caption{Topology example for 36 nodes with 10\% of nodes behaving as attackers: there is one SDN controller, two sinks, and three attackers. The green circle represents the radio range of all nodes. positions for 36 nodes when 10\% of nodes are attackers}
\label{fig:36_nodes_attackers}
\end{center}
\end{figure}

\begin{figure}[t]
\begin{center}
\subfloat[FDFF attack\label{fig:a1_3_100}]{
    \includegraphics[width=0.35\textwidth]{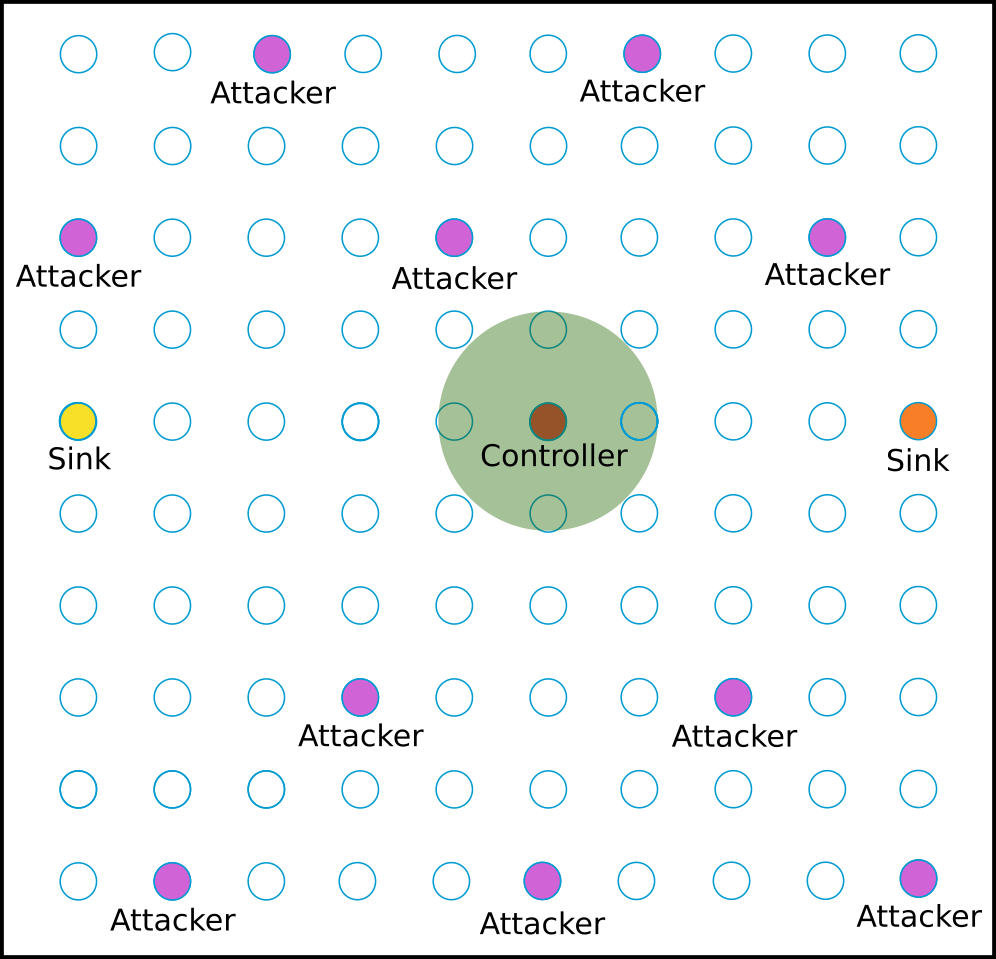}}
\hspace{\fill}
\subfloat[FNI attack\label{fig:a3_3_100}]{
    \includegraphics[width=0.35\textwidth]{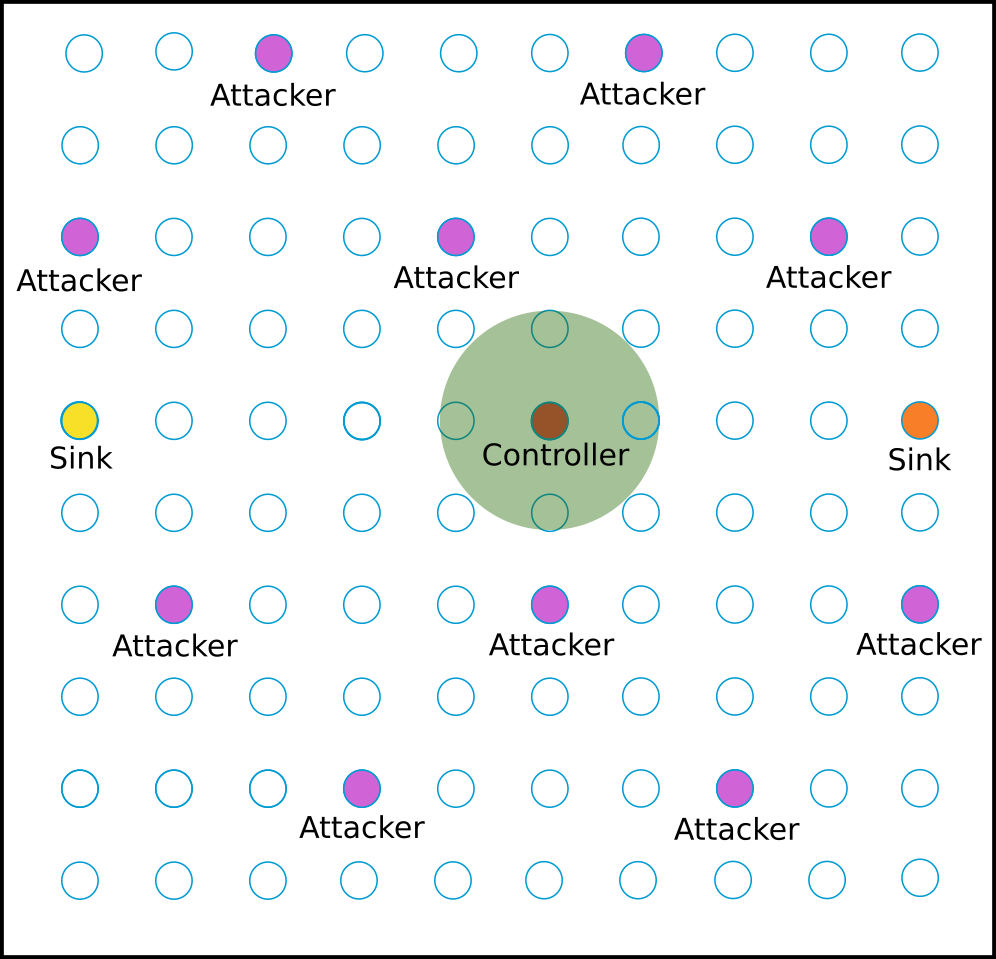}}
\caption{Topology example for 100 nodes with 10\% of nodes behaving as attackers: there is one SDN controller, two sinks, and ten attackers. The green circle represents the radio range of all nodes.}
\label{fig:100_nodes_attackers}
\end{center}
\end{figure}

The SDWSN implementation uses IT-SDN, without changing the default configuration \cite{8805072}, and the simulations were performed using COOJA simulator \cite{Osterlind2006}, emulating Tmote sky motes. We used fully bidirectional square grid topologies with 36 and 100 nodes, one controller, two sinks: one sink to receive data packets and the other one to receive management packets. The controller was in the center of the grid and the sinks were in the middle of the grid edge, since this location gave a better performance in terms of delay, control overhead, energy consumption, and delivery rate according to \cite{8805072}. The attackers were distributed into the network semi-randomly under the condition that two or more attackers can not be neighbors and this distribution remains equal on every scenario replication. Figs. \ref{fig:36_nodes_attackers} and \ref{fig:100_nodes_attackers} show the attackers distribution for 36 nodes and 100 nodes, respectively, when 10\% of nodes are attackers. The green circle around the controller represents the devices' radio range. 

The sensor nodes were programmed to transmit one data packet every 30 seconds and one management packet every 2 minutes, both with payload of 10 bytes. The data packets contained the application information and the management packets contained the information required by the network management plane \cite{Thamires}.
The data packets delivery rate and the control packets overhead were observed every two minutes, considering the exchange of messages in the whole network during this window of time. The delivery rate was calculated by dividing the number of data packets successfully received by the number of data packets sent. The control packets overhead was quantified as the number of control packets sent. Since we took samples every two minutes, we decided to run each single simulation for 10 hours. During the first 8 hours the network operated normally (i.e., for $240$ samples there was no change), then the attack was triggered. This imposed a bound  $m<240$. Table \ref{tab:parameters} summarizes the simulation's and IT-SDN's most important parameters.

\begin{table}[htb]
\renewcommand{\arraystretch}{1.3}
% if using array.sty, it might be a good idea to tweak the value of
% \extrarowheight as needed to properly center the text within the cells
% \todo{adicionar novos cenários (adicionado tipos de flow setup e P2P-RPL nos MOPs)} 
% \todo{olhar tabela II de \cite{sdnwise-testbed} e ver o que eles colocaram como referência}
\caption{Simulation Parameters}
\label{tab:parameters}
\centering
% \resizebox{\columnwidth}{!}{%
\begin{tabular}{ll}
\hline
\multicolumn{2}{|l|}{\textbf{Simulation parameters}}                                                                         \\ \hline
\multicolumn{1}{|l|}{Topology}                          & \multicolumn{1}{l|}{Square grid}                          \\ \hline
\multicolumn{1}{|l|}{Number of nodes}                   & \multicolumn{1}{l|}{36 and 100}     
\\ \hline
\multicolumn{1}{|l|}{Simulation time}                & \multicolumn{1}{l|}{36000 s}     
\\ \hline
\multicolumn{1}{|l|}{Node boot interval}                & \multicolumn{1}{l|}{$[0, 1]$ s}                           \\ \hline
\multicolumn{1}{|l|}{Number of sinks}                   & \multicolumn{1}{l|}{2}                                 \\ \hline
\multicolumn{1}{|l|}{Sinks position}                    & \multicolumn{1}{l|}{Middle of the grid edge}              \\ \hline
\multicolumn{1}{|l|}{controller position}               & \multicolumn{1}{l|}{center}                               \\ \hline
\multicolumn{1}{|l|}{Data traffic rate}                 & \multicolumn{1}{l|}{1 packet every 30 seconds}            \\ \hline
\multicolumn{1}{|l|}{Management traffic rate}           & \multicolumn{1}{l|}{1 packet every two minutes}            \\ \hline
\multicolumn{1}{|l|}{Data payload size}                 & \multicolumn{1}{l|}{10 bytes}                             \\ \hline
\multicolumn{1}{|l|}{Management payload size}           & \multicolumn{1}{l|}{10 bytes}                       \\ \hline
\multicolumn{1}{|l|}{Data traffic start time}           & \multicolumn{1}{l|}{$[2, 3]$ min}                         \\ \hline
\multicolumn{1}{|l|}{Radio module power}                & \multicolumn{1}{l|}{0 dB}                         \\ \hline
\multicolumn{1}{|l|}{Distance between neighbors}        & \multicolumn{1}{l|}{50 m}                         \\ \hline
\multicolumn{1}{|l|}{Attacks begins after}                  & \multicolumn{1}{l|}{28800 s}                         \\ \hline

\\ \hline
\multicolumn{2}{|l|}{\textbf{IT-SDN parameters}}                                                                             \\ \hline
\multicolumn{1}{|l|}{Controller retransmission timeout} & \multicolumn{1}{l|}{60 s}                                 \\ \hline
\multicolumn{1}{|l|}{ND protocol}                       & \multicolumn{1}{l|}{Collect-based}                        \\ \hline
\multicolumn{1}{|l|}{Link metric}                       & \multicolumn{1}{l|}{ETX}                                  \\ \hline
\multicolumn{1}{|l|}{Neighbor report max frequency}     & \multicolumn{1}{l|}{1 packer per minute}                  \\ \hline
\multicolumn{1}{|l|}{CD protocol}                       & \multicolumn{1}{l|}{none}                                 \\ \hline
\multicolumn{1}{|l|}{Flow setup}                        & \multicolumn{1}{l|}{source routed}                        \\ \hline
\multicolumn{1}{|l|}{Route calculation algorithm}       & \multicolumn{1}{l|}{Dijkstra}                             \\ \hline
\multicolumn{1}{|l|}{Route recalculation threshold}     & \multicolumn{1}{l|}{$10\%$}                               \\ \hline
\multicolumn{1}{|l|}{{Flow setup types}}     & \multicolumn{1}{l|}{{regular or source routed}}                               \\ \hline
\multicolumn{1}{|l|}{{Flow table size}}     & \multicolumn{1}{l|}{{10 entries}}                               \\ \hline
\end{tabular}
% }
\end{table}

\subsection{Results analysis}

As explained in Section \ref{sec:method-centralized}, we separated our dataset in two groups, one to determine the values of $m$ and $\gamma$ that maximizes $P_{DS}$ and the other one to evaluate the performance of our proposal using these values. In Section \ref{sec:training} we analyze the results of the training experiments and in Section \ref{sec:centralized-performance} we analyze our proposal performance.

\subsubsection{Optimizing $m$ and $\gamma$} \label{sec:training}

The main objective of the these experiments was to determine the parameters $\{m, \gamma\}$ that could provide the best detection performance based on the metric $P_{DS}$. We calculated the $P_{DS}$ metric for all topologies, attack scenarios and combinations of $m$ and $\gamma$. Then we analyzed the results for $\alpha \in \{0.90, 0.95, 0.99\}$. The first results showed that in $90\%$ of all cases $P_{DS}$ was maximized when $m = 200$, turning this value a universally optimal choice and the $m$ value used for the remaining of the analysis. This means that when running the online detector, no training is required, other than the observation of $200$ samples of normal network operation.

\begin{figure*}[t]
   \subfloat[$A=1$ and $B=0$\label{fig:P_fdff_a1}]{%
      \includegraphics[width=0.31\textwidth]{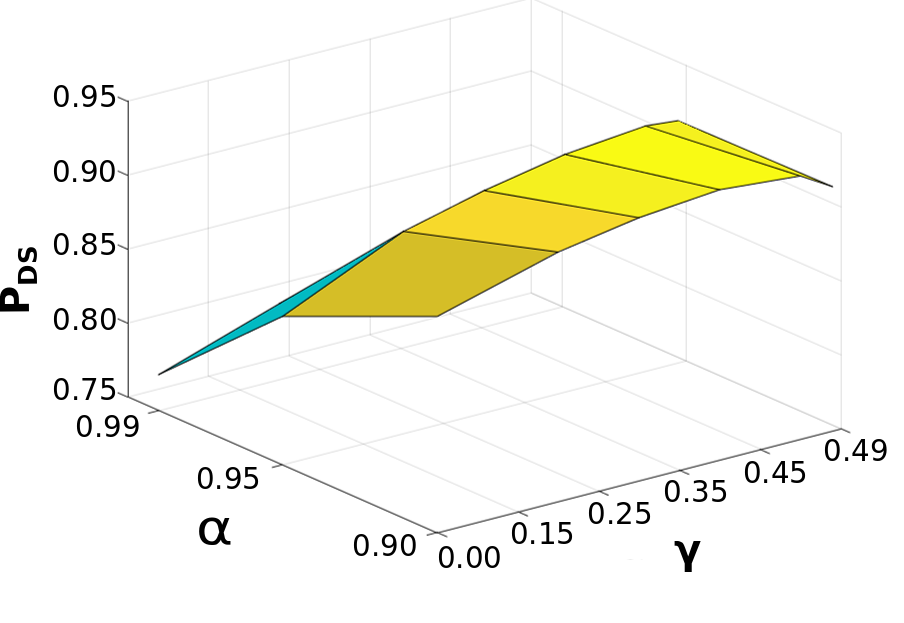}}
\hspace{\fill}
   \subfloat[$A=B=0.5$\label{fig:P_fdff_a05} ]{%
      \includegraphics[width=0.31\textwidth]{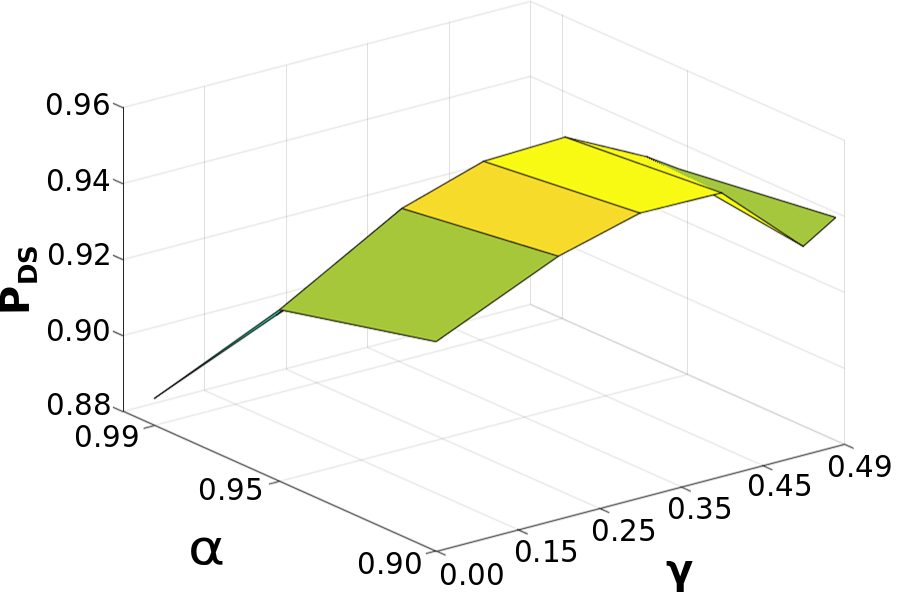}}
\hspace{\fill}
   \subfloat[$A=0$ and $B=1$\label{fig:P_fdff_a0}]{%
      \includegraphics[width=0.31\textwidth]{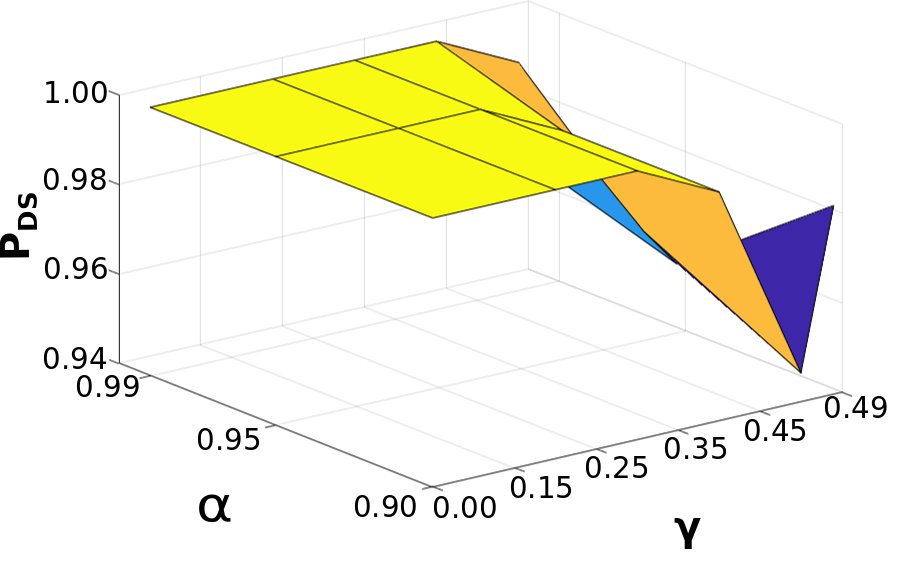}}\\
\caption{Metric $P_{DS}$ in function of $\gamma$ and $\alpha$ for FDFF attack: (a) shows $P_{DS}$ when prioritizing quickest detection, (b) shows $P_{DS}$ when giving the same weight to detection speed and detection rate, and (c) shows $P_{DS}$ when prioritizing detection rate}
\label{fig:P_fdff}
\end{figure*}

For the next part, we separated the results grouping each attack by monitoring metric: for the FDFF attack we analyzed the control overhead CP detection results, and for the FNI attack we analyzed the data packets delivery rate CP detection results, based on the results in \cite{icc-2020}.
Fig. \ref{fig:P_fdff} shows the average value of $P_{DS}$ as a function of $\gamma$ and $\alpha$ for the case of FDFF attack. In Fig. \ref{fig:P_fdff_a1} we observed that in the case of prioritizing faster detection (i.e. $A=1$) the higher results of $P_{DS}$ are for $\gamma = \{0.35, 0.45\}$ and the lower results of $P_{DS}$ are for $\gamma = \{0, 0.15\}$. Opposite, in Fig. \ref{fig:P_fdff_a0} we observed that prioritizing the detection rate, the higher values of $P_{DS}$ are for $\gamma = \{0, 0.15, 0.25\}$, reaching $P_{DS}=1$.

%Lastly, if $A = B = 0.5$, which means we do not prioritize nor faster detection neither detection rate, higher values of $P_{DS}$ are for $\gamma = \{0.25, 0.35\}$ and its curve is flatter than the other two cases. 

\begin{figure*}[htb]
   \subfloat[$A=1$ and $B=0$\label{fig:P_fni_a1}]{%
      \includegraphics[width=0.31\textwidth]{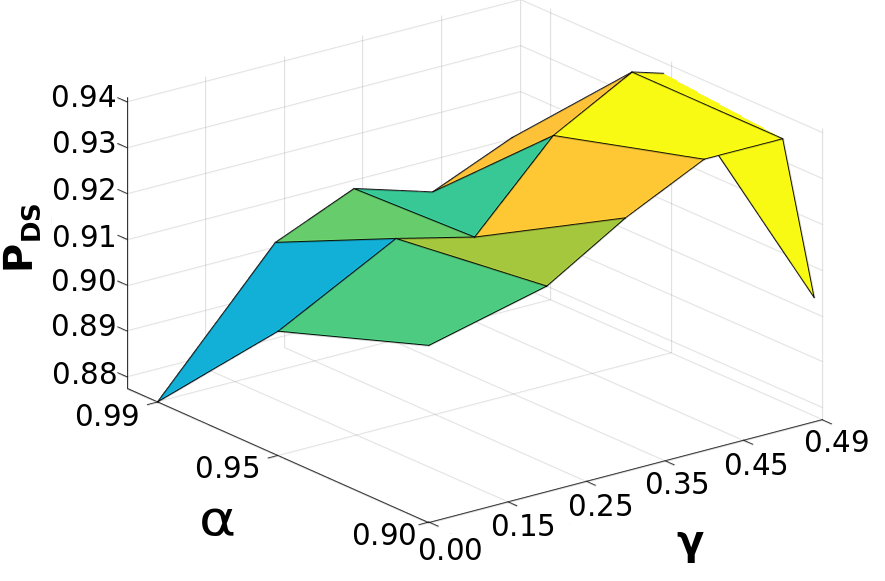}}
\hspace{\fill}
   \subfloat[$A=B=0.5$\label{fig:P_fni_a05} ]{%
      \includegraphics[width=0.31\textwidth]{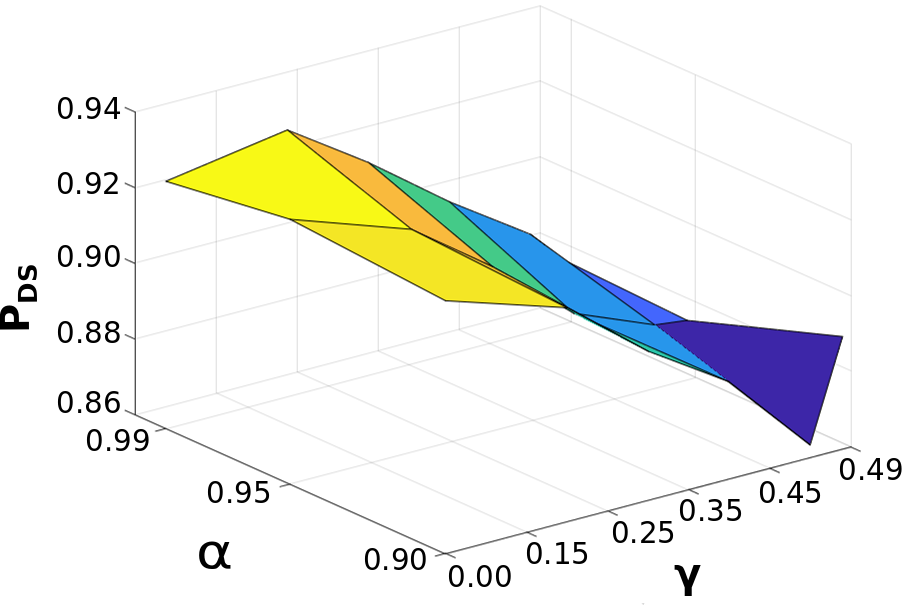}}
\hspace{\fill}
   \subfloat[$A=0$ and $B=1$\label{fig:P_fni_a0}]{%
      \includegraphics[width=0.31\textwidth]{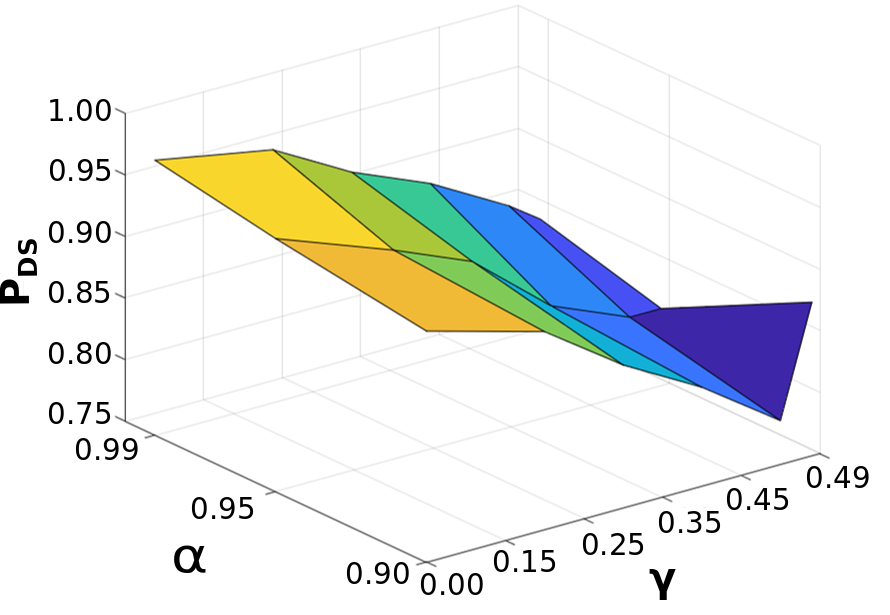}}\\
\caption{Metric $P_{DS}$ in function of $\gamma$ and $\alpha$ for FNI attack: (a) shows $P_{DS}$ when prioritizing quickest detection, (b) shows $P_{DS}$ when giving the same weight to detection speed and detection rate, and (c) shows $P_{DS}$ when prioritizing detection rate}
\label{fig:fnipds}
\end{figure*}

 Fig. \ref{fig:fnipds} shows the average value of $P_{DS}$ for the case of FNI attack. Opposite to the results in Fig. \ref{fig:P_fdff}, in this case they were not as clear-cut as the case for the FDFF attack because lower values of $\gamma$ maximized $P_{DS}$ when $A = B = 0.5$ and $B = 1$, which means the detection rate component has more influence on $P_{DS}$ than the detection speed component. 
 
 %This is also supported by the results in \ref{fig:P_fni_a1} where we observe that considering only the detection speed component (e.g. $A = 1$) the curve is flatter, varying from $0.88$ to $0.91$ when $\alpha = 0.99$ and varying from $0.90$ to $0.93$ when $\alpha = \{0.95, 0.90\}$. Conversely, the results in \ref{fig:P_fni_a1}, considering only the detection rate component, $P_{DS}$ varies from $0.75$ to $0.95$ approximately.

From these results we infered that varying $\gamma$ we are able to configure our detector to prioritize faster detection or accuracy. On the other hand, the response is different for both attacks. 
%In Summary, the detection speed component in the FDFF attack detector is more sensible to $\gamma$ than the detection rate component, while in the FNI attack detector we have the opposite scenario. 
In Table \ref{tab:best_gamma} we show the values of $\gamma$ that maximized $P_{DS}$. In cases where more than one value provided the same or very comparable results, we chose one of them arbitrarily.

\begin{table}[t]
\centering
\renewcommand{\arraystretch}{1.2}
\caption{$\gamma$ that maximizes $P_{DS}$}
\label{tab:best_gamma}
\begin{tabular}{lccc}
\hline
\multicolumn{1}{|c|}{\multirow{2}{*}{\pmb{$P_{DS}$}}} & \multicolumn{3}{c|}{\pmb{$\gamma$}}                                                                                \\ 
\cline{2-4} 
\multicolumn{1}{|c|}{}                             & \multicolumn{1}{c|}{\textbf{$\alpha = 0.90$}} & \multicolumn{1}{c|}{\textbf{$\alpha = 0.95$}} & \multicolumn{1}{c|}{\textbf{$\alpha = 0.99$}} \\
% \hline                                                   & \multicolumn{1}{l}{}               & \multicolumn{1}{l}{}               & \multicolumn{1}{l}{}               \\ 
\hline
\multicolumn{4}{|c|}{\textbf{Best $\gamma$ for control overhead CP detector}}                                                                                                         \\ \hline
\multicolumn{1}{|l|}{$A=1$ and $B=0$}                  & \multicolumn{1}{c|}{0.45}          & \multicolumn{1}{c|}{0.45}          & \multicolumn{1}{c|}{0.45}          \\ \hline
\multicolumn{1}{|l|}{$A=0.8$ and $B=0.2$}              & \multicolumn{1}{c|}{0.35}          & \multicolumn{1}{c|}{0.35}          & \multicolumn{1}{c|}{0.45}          \\ \hline
\multicolumn{1}{|l|}{$A=0.5$ and $B=0.5$}              & \multicolumn{1}{c|}{0.25}          & \multicolumn{1}{c|}{0.35}          & \multicolumn{1}{c|}{0.45}          \\ \hline
\multicolumn{1}{|l|}{$A=0.2$ and $B=0.8$}              & \multicolumn{1}{c|}{0.25}          & \multicolumn{1}{c|}{0.25}          & \multicolumn{1}{c|}{0.35}          \\ \hline
\multicolumn{1}{|l|}{$A=0$ and $B=1$}                 & \multicolumn{1}{c|}{0}             & \multicolumn{1}{c|}{0}             & \multicolumn{1}{c|}{0}             \\ 
% \hline                                                   & \multicolumn{1}{l}{}               & \multicolumn{1}{l}{}               & \multicolumn{1}{l}{}               \\
\hline
\multicolumn{4}{|c|}{\textbf{Best $\gamma$ for delivery rate CP detector}}                                                                                                          \\ \hline
\multicolumn{1}{|l|}{$A=1$ and $B=0$}                  & \multicolumn{1}{c|}{0.45}          & \multicolumn{1}{c|}{0.45}          & \multicolumn{1}{c|}{0.45}          \\ \hline
\multicolumn{1}{|l|}{$A=0.8$ and $B=0.2$}              & \multicolumn{1}{c|}{0}             & \multicolumn{1}{c|}{0.15}          & \multicolumn{1}{c|}{0.15}          \\ \hline
\multicolumn{1}{|l|}{$A=0.5$ and $B=0.5$}              & \multicolumn{1}{c|}{0}             & \multicolumn{1}{c|}{0}             & \multicolumn{1}{c|}{0.15}          \\ \hline
\multicolumn{1}{|l|}{$A=0.2$ and $B=0.8$}              & \multicolumn{1}{c|}{0}             & \multicolumn{1}{c|}{0}             & \multicolumn{1}{c|}{0}             \\ \hline
\multicolumn{1}{|l|}{$A=0$ and $B=1$}                  & \multicolumn{1}{c|}{0}             & \multicolumn{1}{c|}{0}             & \multicolumn{1}{c|}{0}             \\ \hline
\end{tabular}
\end{table}

\subsubsection{Centralized detector performance} \label{sec:centralized-performance}

\begin{figure}[tb]
    \centering
    \includegraphics[width=0.40\textwidth]{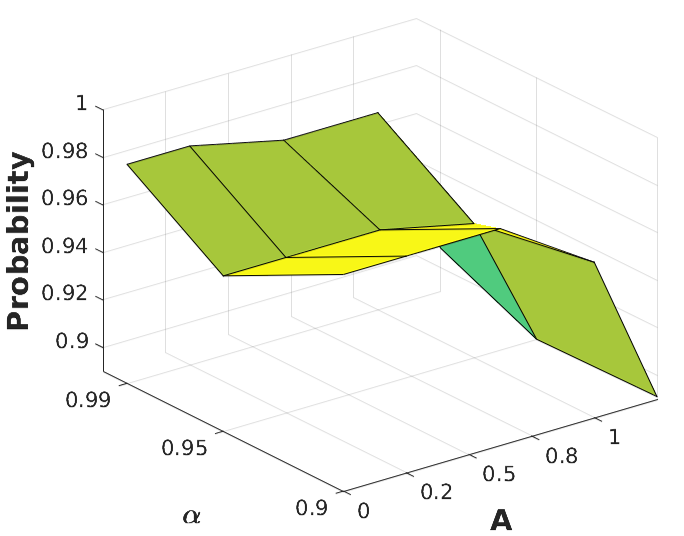}
    \caption{Probability of control overhead CP detector being
triggered first in case of FDFF attack}
    \label{fig:prob_fdff_overhead}
\end{figure}

\begin{figure}[tb]
    \centering
    \includegraphics[width=0.40\textwidth]{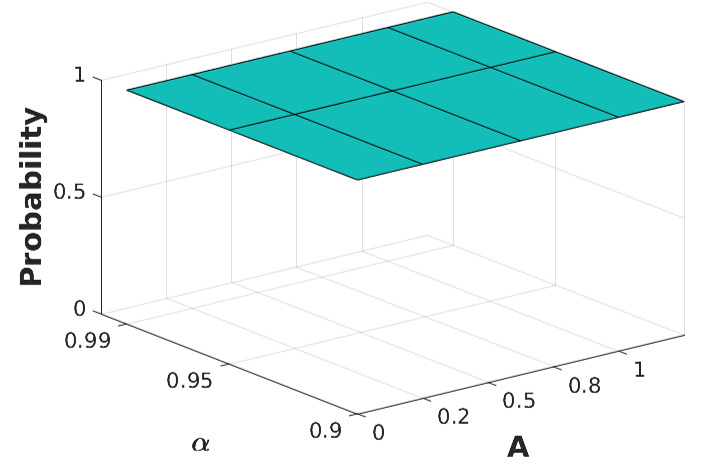}
    \caption{Probability of data packets delivery rate CP detector being
triggered first in case of FNI attack}
    \label{fig:prob_fni_data}
\end{figure}

For this part we set up two detectors running simultaneously %, one monitoring the control overhead and the other one to monitor the data packets delivery rate. Both detectors were configured according to the values in Table \ref{tab:best_gamma} and 
using $m=200$. %according to the decision explained in Section \ref{sec:training}. 
The first experiment was devised to identify the type of the attack based on the first detector triggered. %For this, we counted all the correctly detected attacks and grouped them according to the CP detector triggered. 
Fig. \ref{fig:prob_fdff_overhead} shows the probability of the control overhead CP detector being triggered first in case of FDFF attack. These results showed that in the worst case the detector monitoring the control overhead has a probability between $0.89$ and $0.98 $ of being triggered first in case of FDFF attack. In case of FNI attack, the detector monitoring the data packets delivery rate was triggered first in $100\%$ of the events, as shown in Fig. \ref{fig:prob_fni_data}. These results showed that there is evidence to support the conjecture drawn up in our previous works about the relation metric / attack.  

\begin{figure*}[tb]
   \subfloat[Detection rate\label{fig:DR_FDFF_central}]{%
      \includegraphics[width=0.51\textwidth]{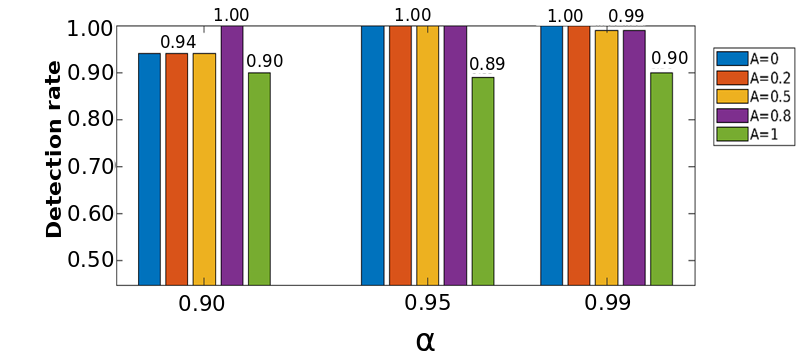}}
\hspace{\fill}
   \subfloat[Detection speed\label{fig:1-s_FDFF_central} ]{%
      \includegraphics[width=0.48\textwidth]{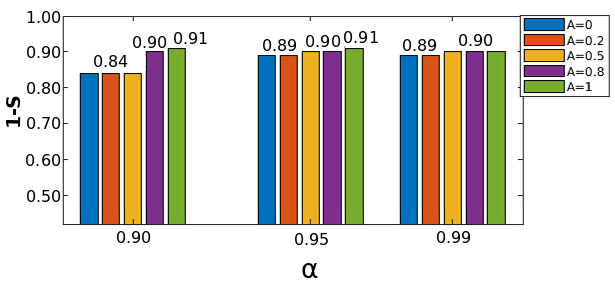}}
\caption{Detection performance of FDFF attack using $\gamma$ and $m$ values that optimize $P_{DS}$ for five different cases: $\{A, B\} = \{\{1, 0\}, \{0.8, 0.2\}, \{0.5, 0.5\}, \{0.2, 0.8\}, \{0, 1\}\}$}
\label{fig:FDFF_central}
\end{figure*}

Next we analyze the detection performance using the parameters that maximize $P_{DS}$. Fig. \ref{fig:FDFF_central} depicts the detection rate $DR$ and the metric $1-S$ when the network is under FDFF attack. %In terms of detection rate, four out of five configurations achieved $DR \geq \alpha$. In terms of detection speed, $1-S \geq 0.90$ for $A = \{0.8, 1\}$. 
Considering both $DR$ and $1-S$ the results for $A = 0.8$ provided the best trade off. 

\begin{figure*}[tb]
   \subfloat[Detection rate\label{fig:DR_FNI_central}]{%
      \includegraphics[width=0.49\textwidth]{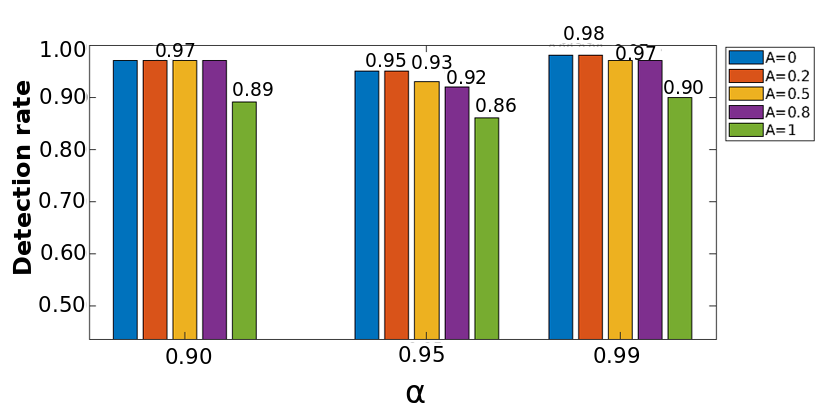}}
\hspace{\fill}
   \subfloat[Detection speed\label{fig:1-s_FNI_central} ]{%
      \includegraphics[width=0.51\textwidth]{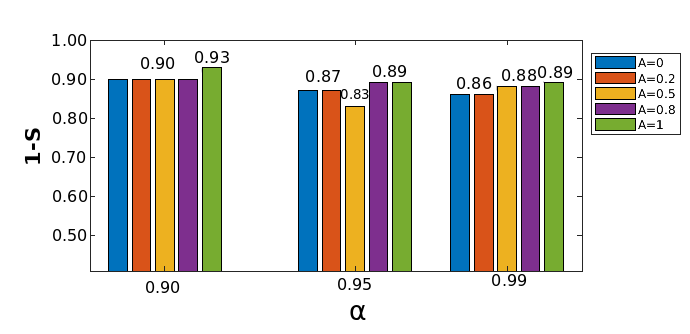}}
\caption{Detection performance of FNI attack using $\gamma$ and $m$ values that optimize $P_{DS}$ for five different cases: $\{A, B\} = \{\{1, 0\}, \{0.8, 0.2\}, \{0.5, 0.5\}, \{0.2, 0.8\}, \{0, 1\}\}$}
\label{fig:FNI_central}
\end{figure*}

Fig. \ref{fig:FNI_central} shows the detection rate and the detection speed metrics for the FNI attack using the identified values of $\gamma$. 
%In this case, four out of five configurations achieved $DR \geq \alpha$ for $\alpha = 0.90$; two out of five configurations achieved $DR \geq \alpha$ for $\alpha = 0.95$; and zero configurations achieved $DR \geq \alpha$ for $\alpha = 0.99$. 
In terms of detection speed, $A=0$ obtained the fastest detection, as intuitively expected based on the results from Fig. \ref{fig:fnipds}. Comparing the results for $A=1$ and $A=0$, we can maximize $DR$ at the cost of $0.03$ in $1-S$, which is equivalent to $1.8$ samples. On the other hand, if we are looking for fastest detection, $DR$ drops to $0.90$ or below.

\begin{figure*}[tb]
   \subfloat[Detection rate\label{fig:DR_total_central}]{%
      \includegraphics[width=0.49\textwidth]{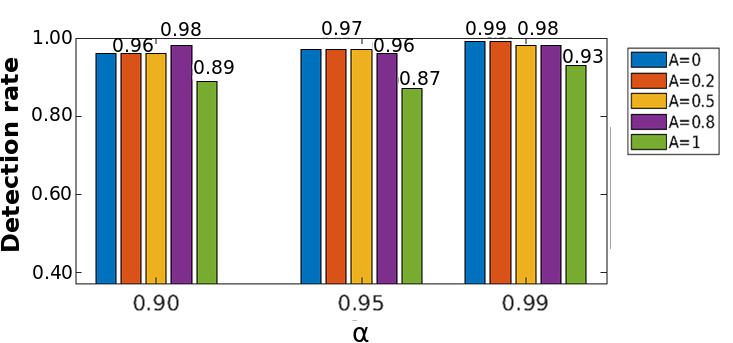}}
\hspace{\fill}
   \subfloat[Detection speed\label{fig:1-s_total_central} ]{%
      \includegraphics[width=0.50\textwidth]{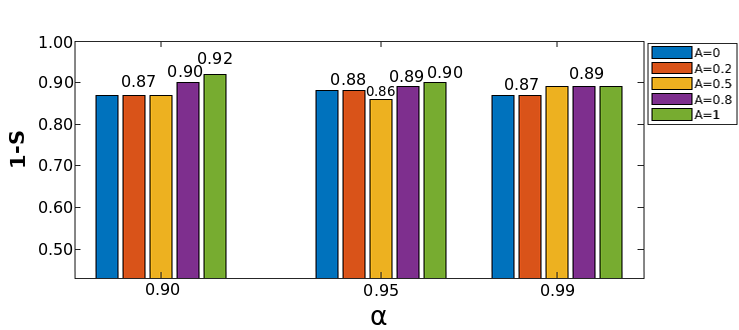}}
\caption{Detection performance of FDFF and FNI attacks using $\gamma$ and $m$ values that optimize $P_{DS}$ for five different cases: $\{A, B\} = \{\{1, 0\}, \{0.8, 0.2\}, \{0.5, 0.5\}, \{0.2, 0.8\}, \{0, 1\}\}$}
\label{fig:total_central}
\end{figure*}

The last scenario we analyzed was the detection performance irrespective of the type of the attack. In this case both detectors were running simultaneously in a network prone to both FDFF and FNI attacks. The results in Fig. \ref{fig:total_central} showed a detection rate over $\alpha$ when $A={0, 0.2, 0.5, 0.8}$ for $\alpha = {0.90, 0.95}$. When $\alpha = 0.99$, $DR = \alpha$ for $A = {0, 0.2}$ only. This means, if we want to maximize the detection rate we need to use the configuration for $A = {0, 0.2}$. In terms of detection speed, as shown in Fig. \ref{fig:1-s_total_central}, to maximize the detection rate means a lag of 3 samples in average with respect to the fastest detection result obtained.   

Summarizing Section \ref{sec:centralized-approach}, we split our dataset in two sets: one for identifying the optimal values of $m, \gamma$ and the other one for validation. We chose the pairs $(m, \gamma)$ that maximized the detection performance metric $P_{DS}$ based on the results from the experiments on the training dataset. Our results showed that in 90\% of cases $m = 200$ maximized the metric $P_{DS}$. With respect to $\gamma$, we observed that using $\gamma={0.45, 049}$ we reduced the time to detect the attack but this had an adverse effect on the detection rate. Conversely, when $\gamma={0, 0.15}$ we maximized the detection at the cost of delaying the detection. Then, we tested the CP detectors on the validation dataset using the parameters values chosen before. Results showed that we were able to detect the attack with $DR \geq \alpha$ when $B > A$. On the other hand, if we prioritize fastest detection, the detection rate drops to $0.93$ or below. 
%Lastly, we showed that there is a probability over $0.89$ that in case the control overhead CP detector is triggered before the data packets delivery rate, the network is under FDFF attack. Moreover, when the network was under an FNI attack, the detector monitoring the data packets delivery rate always raised the alarm first. 
In conclusion, we provided concrete evidence to support the relation between monitored metric and the type of attack.

\section{Distributed Detection} \label{sec:distributed-approach}

In this section we explain our distributed detection proposal for DoS attacks in SDWSN. The central idea is to implement one CP detector on individual nodes (potentially on every node). To the best of our knowledge, intrusion detection at the individual sensor level breaks new ground. In case of a CP detected, the sensor warns the controller about it (which in turn sends this information to the security application through the Network Services Abstraction Layer and the security application decides whether the network is under attack or not). 

Our goal is to investigate whether the detection of FDFF and FNI attacks is feasible on individual nodes. Our hypothesis is that it could be possible if metrics related to the number of control packets exchange and the active state time (i.e., the time the node is not on sleeping mode) are monitored. To test our hypothesis, we run the CP detector on every node and monitored the following metrics: the processing time, the transmitting time, the number of control packets received, and the number of control packets transmitted. The processing time is the time the node remains with the microprocessor in active state and the transmitting time is the time the node remains with the radio module turned on transmitting packets. In the experiments presented below based on Contiki 3.0, both metrics can be obtained using Energest \cite{energest}, a tool to monitor device's hardware usage. Furthermore, the number of control packets received or transmitted can be obtained by programming every node to print every packet sent and received and using COOJA simulator's serial output this information can be copied in a text document. 
%We chose these metrics because, as explained in Section \ref{sec:problem}, the attacks we implemented increased the processing overhead on the nodes close to the attacker and increased the control packets overhead in the network.
% 
\subsection{Experimental setup and results}
\label{sec:results_distributed}
We used a dataset of 120 simulations divided in two groups: half for the FDFF attack and the other half for the FNI attack. For both attacks we simulated grid topologies of 36 and 100 nodes where $10\%$ of nodes were attackers. For these experiments we prioritized detection accuracy over detection speed, thus we configured the detector using $\gamma = 0$ and set the target $\alpha=0.99$. In the case of the monitoring period of no change, we set $m = 200$ according to the results obtained in Section \ref{sec:centralized-approach} to maximize the detection performance.

%We analyzed the CP detection on every node when the network was under attack using four metrics: processing time, transmitting time, the number of control packets received, and the number of control packets transmitted. 

We evaluated the detection performance on every node monitoring each metric separately, i.e., running only one detector at time due to memory constraints on the nodes. For this evaluation we calculated the detection probability of every node on each scenario. 
We maintained the same simulation parameters and attackers positions used for the centralized detection experiments. The parameters are summarized in Table \ref{tab:parameters} and the attackers position are represented in Figs. \ref{fig:36_nodes_attackers} and \ref{fig:100_nodes_attackers}.
%\subsection{Results and discussion for the distributed detection} \label{sec:results_distributed}
% Comparison of centralized and distributed? Results combined
% Centralized -> Distributed 2 phase.
Our detection performance analysis is based on three perspectives under the condition the network is under attack: (i) probability of CP detection on each node; (ii) percentage of nodes reporting with high detection rates; (iii), and location of nodes reporting high detection rates.% For this analysis we calculated the CP detection probability on every node. Since we prioritized the detection probability $P_{DR}$, we configured the detector using $\gamma = 0$ targeting $\alpha=0.99$. 

\subsubsection{Results for FDFF attack}

\begin{figure*}[t]
  \subfloat[36 nodes\label{fig:density_fdff_36}]{%
      \includegraphics[width=0.49\textwidth]{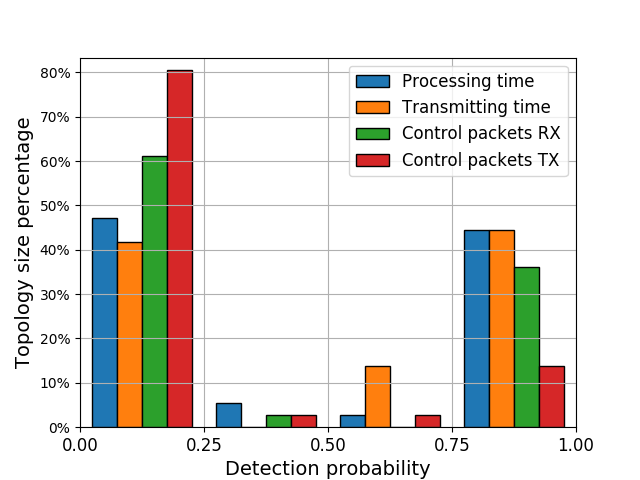}}
\hspace{\fill}
  \subfloat[100 nodes\label{fig:density_fdff_100} ]{%
      \includegraphics[width=0.50\textwidth]{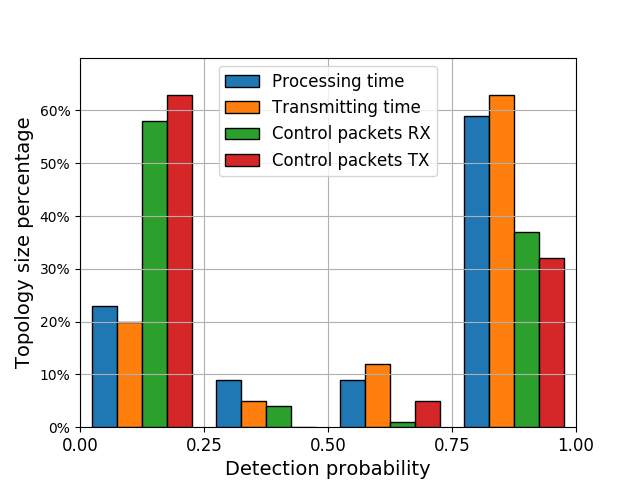}}
\caption{Detection probability distribution of FDFF attack: Comparison of detection probability when monitoring the processing time, transmitting time, control packets received, and control packets transmitted. The ``x'' axis represents the detection probability divided in four groups: $[0, 0.25)$, $[0.25, 0.50)$, $[0.50, 0.75])$, and $[0.75, 1]$. The ``y'' axis represents the percentage of the total nodes that obtained this detection probability. (a) shows the results for 36 nodes and (b) shows the results for 100 nodes.}
\label{fig:density_fdff}
\end{figure*}

Fig. \ref{fig:density_fdff} shows the detection probability distribution when the network is under FDFF attack for 36 and 100 nodes. This means, the percentage of the total number of nodes with very low ($0 \leq P_{DR} \leq 0.25$), low ($0.25 < P_{DR} \leq 0.50$),  high ($0.50 < P_{DR} \leq 0.75$), or very high ($0.75 < P_{DR} \leq 1.00$) detection rates. In the case of 36 nodes, as shown in Fig. \ref{fig:density_fdff_36}, %we have similar percentage of nodes with very low detection rate and with very high detection rate when monitoring the processing time and the transmitting time. From this, 
we noticed there is a large percentage of nodes that have a very high detection rate for FDFF attacks when monitoring the processing time or the transmitting time. 
%The results monitoring the control packets received and transmitted showed peaks in low and high detection probabilities, but with a higher percentage of nodes in the case of low detection probability.   
The results in the case of 100 nodes (Fig. \ref{fig:density_fdff_100}) showed as well that for time based metrics a large portion of the network will identify with very high detection rates the attacks.

\begin{figure*}[t]
  \subfloat[36 nodes\label{fig:hist_fdff_36}]{%
      \includegraphics[width=0.49\textwidth]{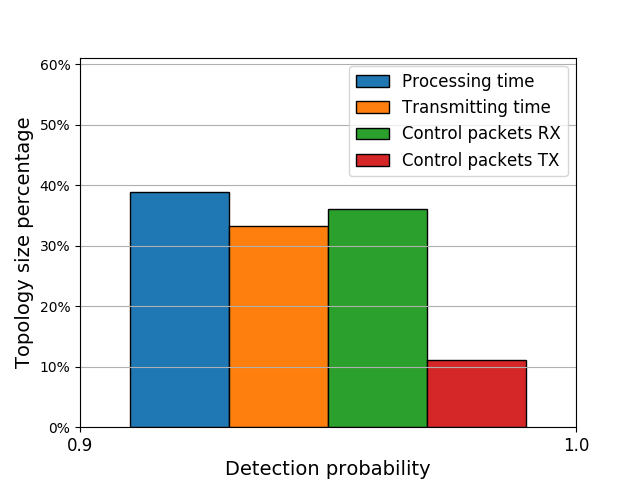}}
\hspace{\fill}
  \subfloat[100 nodes\label{fig:hist_fdff_100} ]{%
      \includegraphics[width=0.50\textwidth]{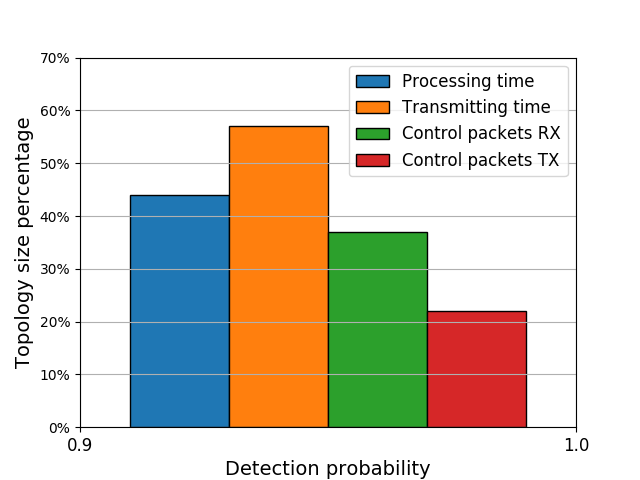}}
\caption{Percentage of nodes with detection probability larger than $90\%$ for the FDFF attack: Comparison of detection probability when monitoring the processing time, transmitting time, control packets received, and control packets transmitted. The ``y'' axis represents the percentage of the total nodes with high detection probability. (a) shows the results for 36 nodes and (b) shows the results for 100 nodes.}
\label{fig:hist_fdff}
\end{figure*}

%The detection probability distribution results when the network was under an FDFF attack showed that monitoring any of the four metrics, there is a peak of percentage of nodes with high detection probabilities. 
Next, we further zoomed in detection probabilities greater than $0.90$, shown in %, since we defined before (Section \ref{sec:relatedwork}) this value to determine which related works obtained high detection rate or not. 
Fig. \ref{fig:hist_fdff}. % shows the percentage of nodes with a $P_{DR}\geq 0.90$ 
when the network is under FDFF attack for topologies with 36 (Fig. \ref{fig:hist_fdff_36}) and 100 (Fig. \ref{fig:hist_fdff_100}) nodes. For the case of 36 nodes, when monitoring the control packets transmitted around $12\%$ of nodes reported an alarm in at least $90\%$ of times the network was under an FDFF attack. On the other hand, the percentage of nodes reporting an alarm increased to over $33\%$ when monitoring either the processing time, the transmitting time, or the control packets received. %This means, in the hypothetical case where we are running an intrusion detection policy based on the number of alarms received by the controller, we reduced the detection chances if monitoring the control packets transmitted instead of any of the other three metrics.

The percentage of nodes reporting an alarm increases in general with the network size, obtaining the highest result when monitoring the transmitting time and the lowest result when monitoring the control packets transmitted. %Opposite to the results for 36 nodes, where the percentage of nodes reporting an alarm was similar when monitoring either the processing time, the transmitting time, or the control packets received, for 100 nodes there was a clear difference among the results for all the metrics. The percentage of nodes when monitoring the transmitting time was $12\%$ over the results when monitoring the processing time, and $20\%$ over the results when monitoring the control packets received. 
In brief, for 36 nodes the percentage of nodes reporting alarms with $P_{DR} \geq 0.90$ was similar when monitoring either the processing time, transmitting time, or control packets received. However, for 100 nodes the results when monitoring the transmitting time were clearly over the results when monitoring any of the other metrics. In the hypothetical case where the nodes have resources to monitor only one metric, the transmitting time is the one that provides the best trade off in terms of percentage of nodes reporting an alarm.   

\begin{figure*}[t]
   \subfloat[Transmitting time\label{fig:heatmap_36_fdff_tx}]{%
      \includegraphics[width=0.49\textwidth]{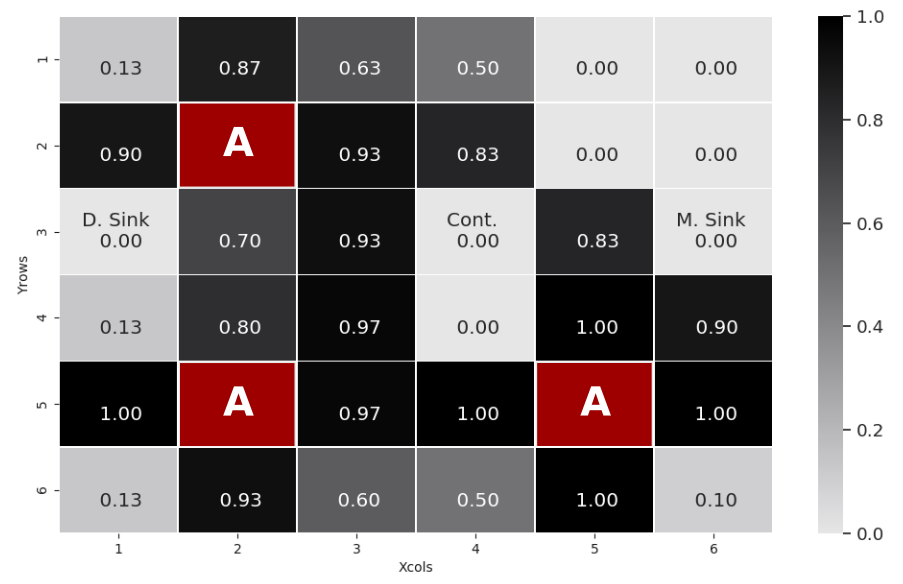}}
\hspace{\fill}
   \subfloat[Control packets received\label{fig:heatmap_36_fdff_rx_ctrl} ]{%
      \includegraphics[width=0.50\textwidth]{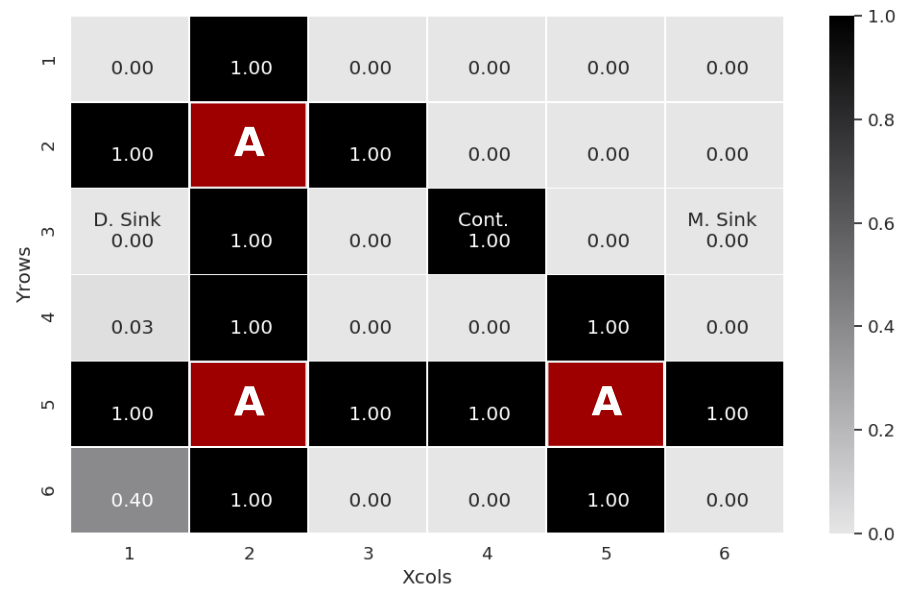}}
\caption{Detection probability heat maps for 36 nodes when the network is under FDFF attack. Each square represents a node in the network and the number inside them is the detection probability result. The red squares with an ``A'' inside are the attackers. (a) shows the results when monitoring the transmitting time and (b) shows the results when monitoring the control packets received.}
\label{fig:heatmap_36_fdff}
\end{figure*}

\begin{figure*}[t]
   \subfloat[Transmitting time\label{fig:heatmap_100_fdff_tx}]{%
      \includegraphics[width=0.49\textwidth]{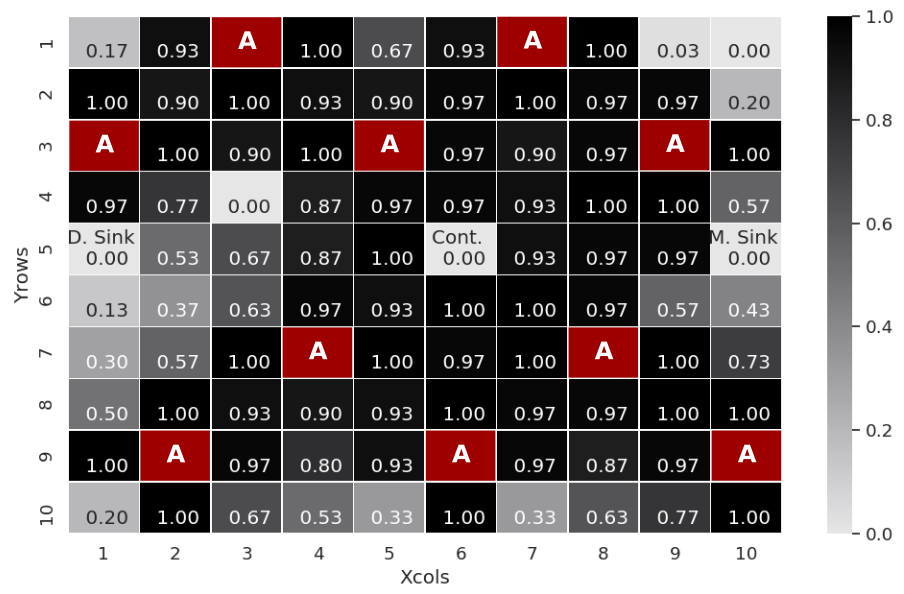}}
\hspace{\fill}
   \subfloat[Control packets received\label{fig:heatmap_100_fdff_rx_ctrl} ]{%
      \includegraphics[width=0.50\textwidth]{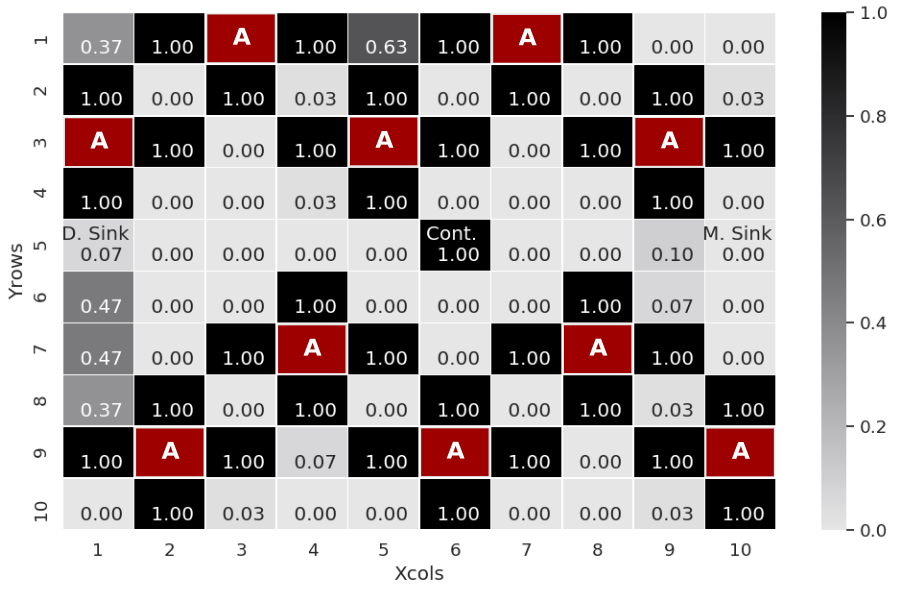}}
\caption{Detection probability heat maps for 100 nodes when the network is under FDFF attack. Each square represents a node in the network and the number inside them is the detection probability result. The red squares with an ``A'' inside are the attackers. (a) shows the results when monitoring the transmitting time and (b) shows the results when monitoring the control packets received.}
\label{fig:heatmap_100_fdff}
\end{figure*}

Subsequently, we analyzed the position of nodes in the network and their detection probabilities. For this analysis we chose the time based metric and the control packets based metric with better detection. Fig. \ref{fig:heatmap_36_fdff} shows the heat maps for 36 nodes when monitoring the transmitting time and the control packets received. From these results we make two observations: i) in the case monitoring the transmitting time, the neighbors of the attackers had higher detection probability than nodes farther; and ii) in the case monitoring the control packets received, excluding the controller and the node on the lower left corner, all nodes reporting an alarm were in the attacker's neighborhood and had a $P_{DR}=1$. For 100 nodes we observed a similar behavior when monitoring the control packets received (Fig. \ref{fig:heatmap_100_fdff_rx_ctrl}), but when monitoring the transmitting time (Fig. \ref{fig:heatmap_100_fdff_tx}) we observed that high detection probability is not exclusive for attackers' neighbors and it is spread all over the topology. This happened because when the network was under attack, the number of control packets increased and this impacted the radio usage of all nodes forwarding those packets. On the other hand, the control packets received is a metric that impacts only the node that receives the packet. In Section \ref{sec:tracking} we explore how to use node's location and address to identify the attackers.

\subsubsection{FNI} \label{sec:fni_prob}

\begin{figure*}[htb]
  \subfloat[36 nodes\label{fig:density_fni_36}]{%
      \includegraphics[width=0.49\textwidth]{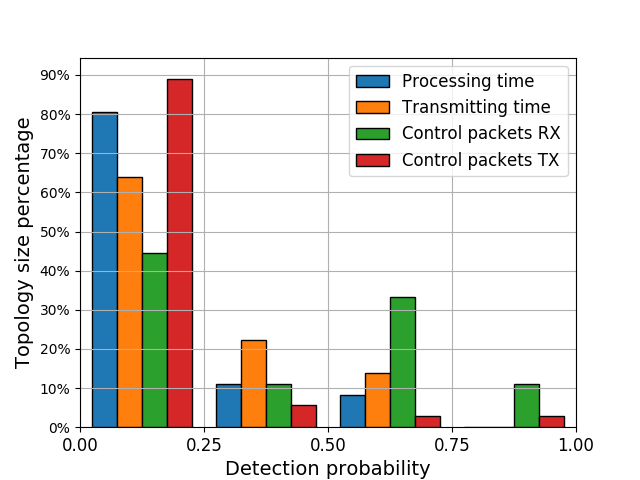}}
\hspace{\fill}
  \subfloat[100 nodes\label{fig:density_fni_100} ]{%
      \includegraphics[width=0.50\textwidth]{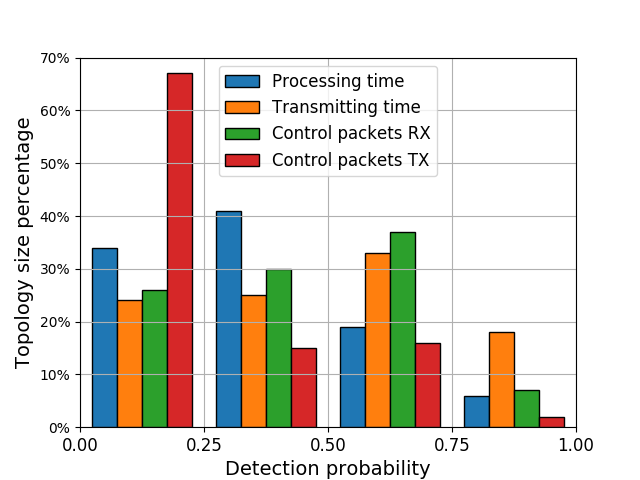}}
\caption{Detection probability distribution of FNI attack: Comparison of detection probability when monitoring the processing time, transmitting time, control packets received, and control packets transmitted. The ``x'' axis represents the detection probability divided in four groups: $[0, 0.25)$, $[0.25, 0.50)$, $[0.50, 0.75])$, and $[0.75, 1]$. The ``y'' axis represents the percentage of the total nodes that obtained this detection probability. (a) shows the results for 36 nodes and (b) shows the results for 100 nodes.}
\label{fig:density_fni}
\end{figure*}

Fig. \ref{fig:density_fni} shows the detection probability density distribution results when the network was under an FNI attack. For 36 nodes (Fig. \ref{fig:density_fni_36}) we observed a similar behavior for all four metrics: high density in probabilities around $0$ and $0.20$ that decreased as the detection probability grew, being the result for control packets received the one with highest density in probabilities over $0.6$. In the case of 100 nodes, the results for control packets transmitted maintained the behavior observed for 36 nodes, with high density in probabilities between $0$ and $0.20$ that decreased for higher probabilities. The results for processing time, transmitting time, and control packets received showed high detection probability density around $0.20$ and $0.50$. Then, for detection probabilities over $0.90$, the highest density was for the transmitting time. The reason why we observed more impact on the transmitting time and the control packets received is because this attack leads to a network reconfiguration using wrong neighborhood information. First, the network reconfiguration means several control packets from the controller to the nodes, which increases this metric on these nodes. Then, since the reconfiguration is based on wrong information, the number of packets retransmission increases, increasing the transmitting time metric as well.   

\begin{figure*}[htb]
  \subfloat[36 nodes\label{fig:hist_fni_36}]{%
      \includegraphics[width=0.49\textwidth]{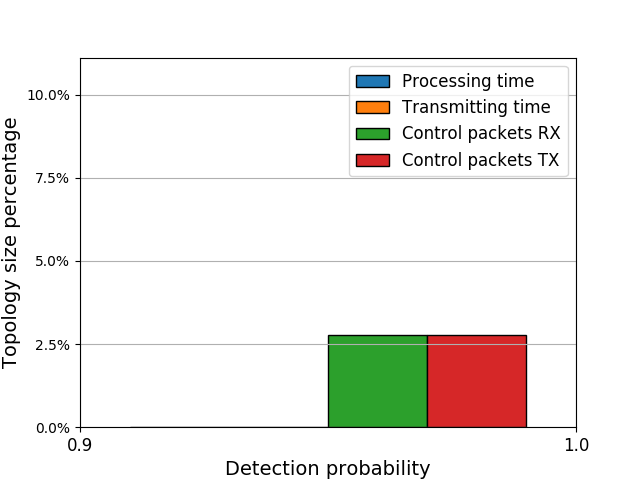}}
\hspace{\fill}
  \subfloat[100 nodes\label{fig:hist_fni_100} ]{%
      \includegraphics[width=0.50\textwidth]{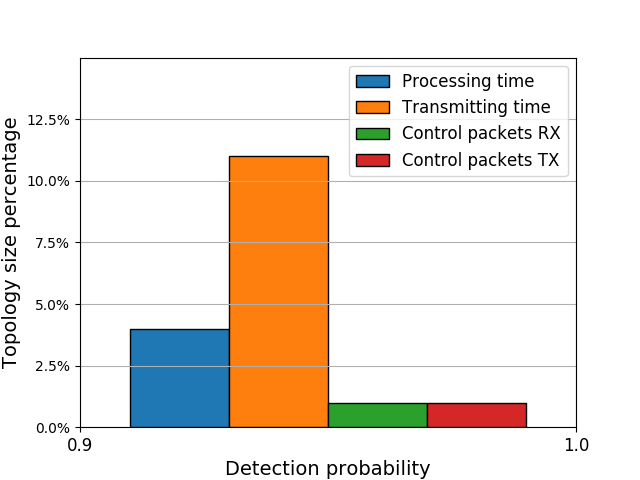}}
\caption{Percentage of nodes with detection probability of FNI attack larger than $90\%$: Comparison of detection probability when monitoring the processing time, transmitting time, control packets received, and control packets transmitted. The ``y'' axis represents the percentage of the total nodes with high detection probability. (a) shows the results for 36 nodes and (b) shows the results for 100 nodes.}
\label{fig:hist_fni}
\end{figure*}

To confirm previous results, we calculated the percentage of nodes reporting an alarm with probabilities $P_{DR} \geq 0.90$. Fig. \ref{fig:hist_fni} shows these results for 36 and 100 nodes. For 36 nodes, $2.7\%$ of nodes obtained a $P_{DR} \geq 0.90$ when monitoring either the control packets received or the control packets transmitted. Since $2.7\%$ represent less than one node, we consider that our distributed proposal is not able to detect an FNI attack in a small topology with a probability above $0.90$. For 100 nodes, the highest result was for the case monitoring the transmitting time, where $11\%$ of nodes obtained a $P_{DR} \geq 0.90$. The percentage of nodes reporting an alarm when monitoring the transmitting time with $P_{DR} \geq 0.90$ is higher for 100 nodes than for 36 nodes because of two reasons: there were more nodes using an attacker to reach the controller, which increased the percentage of nodes affected; and the distance between the attackers and the controller was larger for 100 nodes, which means more nodes participated in the forwarding when doing the network's reconfiguration. For this case, we consider our proposal is able to detect when the network is under an FNI attack with high probability but only when monitoring the transmitting time metric.  

\begin{figure*}[t]
\centering
   \subfloat[36 nodes\label{fig:every_fni_36_rx_ctrl}]{%
      \includegraphics[width=0.49\textwidth]{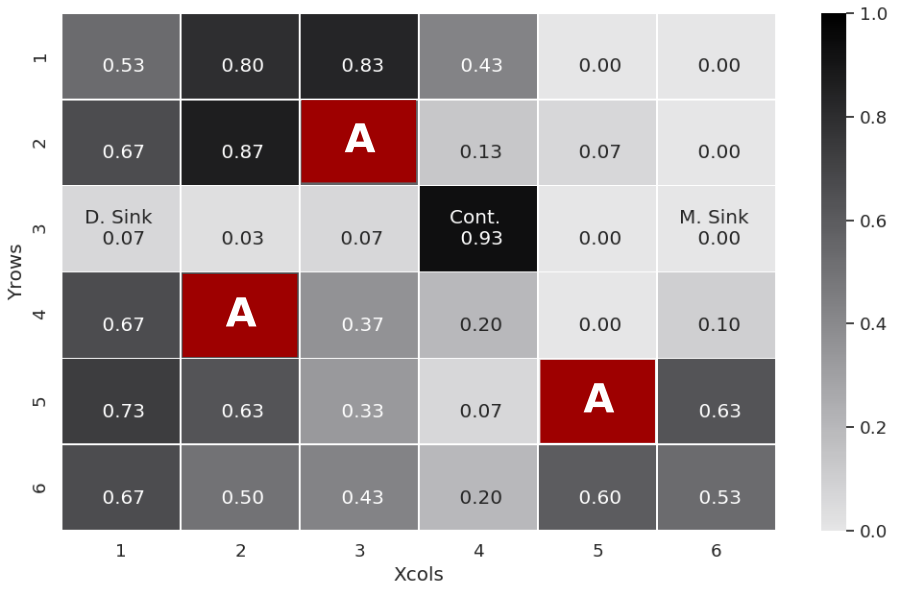}}
\hspace{\fill}
   \subfloat[100 nodes\label{fig:every_fni_100_rx_ctrl} ]{%
      \includegraphics[width=0.50\textwidth]{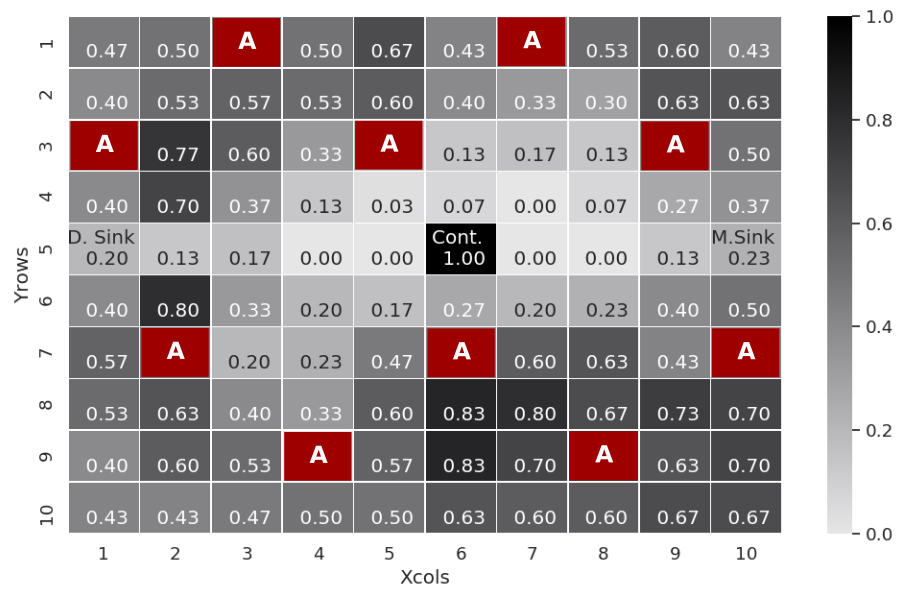}}
\caption{Detection probability heat maps when the network is under FNI attack. Each square represents a node in the network and the number inside them is the detection probability result. The red squares with an ``A'' inside are the attackers. (a) shows the results for 36 nodes and (b) shows the results for 100 nodes.}
\label{fig:every_fni}
\end{figure*}

\begin{figure*}[htb]
\centering
   \subfloat[36 nodes\label{fig:groups_fni_36_rx_ctrl}]{%
      \includegraphics[width=0.49\textwidth]{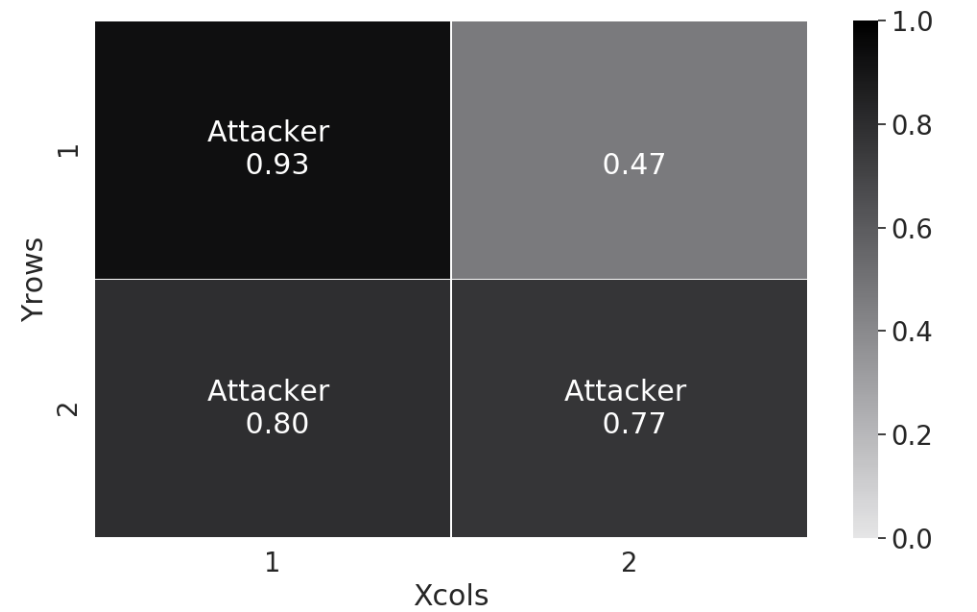}}
\hspace{\fill}
   \subfloat[100 nodes\label{fig:groups_fni_100_rx_ctrl} ]{%
      \includegraphics[width=0.50\textwidth]{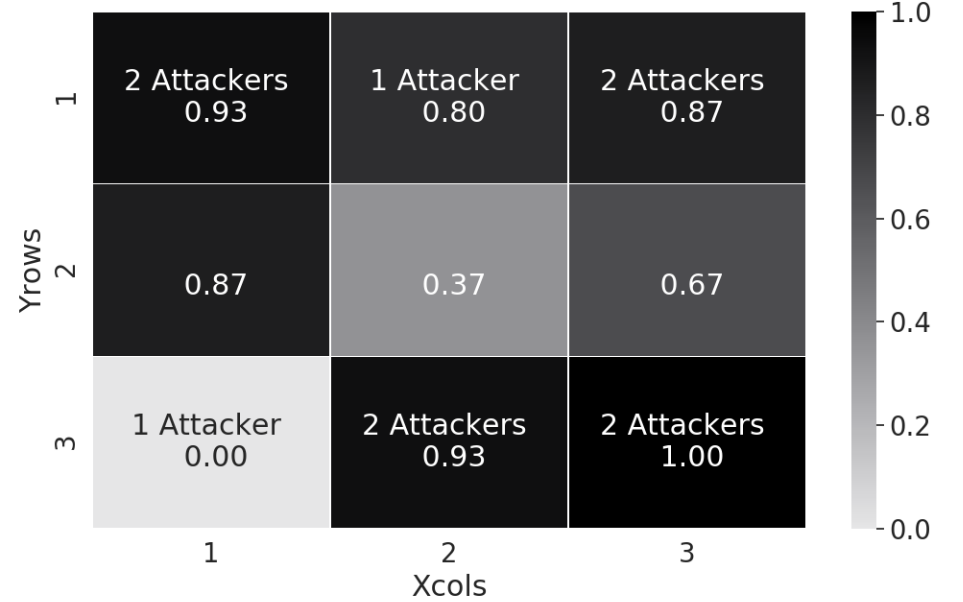}}
\caption{Detection probability heat maps when the network is under FNI attack. Each square represents a group of nodes and the number inside them is the detection probability result aggregating the control packets received information of all nodes in the group. (a) shows the results for 36 nodes and (b) shows the results for 100 nodes. }
\label{fig:groups_fni_36_100}
\end{figure*}

Notwithstanding the detection performance of FNI attack, in Fig. \ref{fig:density_fdff} we observed a high density of nodes reporting alarms in probabilities over $0.50$ when monitoring the control packets received and the transmitting time, thus we decided to investigate the location of those nodes in the topology. We observed that in the cases monitoring the control packets received, as shown in Fig. \ref{fig:every_fni}, some nodes around the attackers concentrated the higher detection probability values, but others also close to the attacker had detection probabilities around zero. The question arises as to why this is observed; the reason being that neighbouring nodes with higher detection probabilities used the attacker to route their packets toward the controller, thus the network misconfiguration reached them first. 
From these results, a second strategy based on data aggregation was motivated, analyzing CP detection per regions (areas). To this end, we divided the 36 nodes in four groups and the 100 nodes in nine groups and created one time series per group. Each sample of this time series represented the sum of time series of all nodes in the group, thus we executed one CP detector per group. Fig. \ref{fig:groups_fni_36_100} shows $P_{DR}$ results for 36 and 100 nodes when monitoring the control packets received. Excluding the groups that contained the controller, in all  cases the detection probability achieved is better than the one obtained by any of the nodes individually. This indicates that with data aggregation we lose granularity but we gain in detection rates.

Summarizing Section \ref{sec:distributed-approach}, we evaluated our CP detection proposal on networks under FDFF and FNI attack, monitoring four metrics obtained from each node: processing time, transmitting time, control packets received, and control packets transmitted. Our results showed in case of FDFF attack, at least $33\%$ of the total of nodes obtained a detection probability equal or over $90\%$ when monitoring the processing time, the transmitting time, or the control packets received. In the cases when the network was under a FNI attack %we had two scenarios: i) for 36 nodes, $2.5\%$ of nodes obtained a detection probability equal or over $90\%$ when monitoring the control packets received and the control packets transmitted; and ii) for 100 nodes we obtained the best result when monitoring the transmitting time, where $11\%$ of nodes reported a detection probability equal or over $90\%$. Since the results for the FNI attack 
were not satisfactory and thus we introduced a second strategy based on data aggregation. Our results showed that using this strategy we increased the detection probability but lost in granularity. 

\section{Attacker detection} \label{sec:tracking}

The results discussed in Sections \ref{sec:results_distributed} and \ref{sec:centralized-approach} showed that the CP detectors for DoS attacks worked for both centralized and distributed detection, but also we observed that the distributed detection provides information that infers the attackers' location. In this direction, our proposal explores the SDN's characteristics by using the controller's global view of the network to identify the attacker's address or location based on the alarms reported by the nodes.

In this section we present and evaluate an algorithm to locate attackers when the network is under an FDFF or FNI attack. We separate our analysis by the type of attack; in subsection \ref{sec:fdff_attacker_detection} we explain and present the results for the FDFF attack and in subsection \ref{sec:fni_attacker_detection} we do the same for the FNI attack.

\subsection{Attacker detection in FDFF attack} \label{sec:fdff_attacker_detection}
Our results in Figs. \ref{fig:heatmap_36_fdff} and \ref{fig:heatmap_100_fdff} showed that when monitoring the control packets the attackers' neighbors had a $P_{DR}=1$, and when monitoring the transmitting time the attackers' neighbors had a $P_{DR}\geq 0.90$. Based on these findings, our proposal is to identify the attackers' IDs based on the alarms reported by their neighbors. To accomplish this, the Security module in the application plane requests neighborhood information to the controller and executes the Algorithm \ref{alg:1}, presented in the following. 

As explained in Algorithm \ref{alg:1}, the Security module waits for an alarm(s) and then requests from the controller the neighborhood information of the nodes reporting. The alarms received are represented by the vector \texttt{alarms\{nodes\}}. Then, the Security module extracts the neighbors of each node in the vector \texttt{alarms\{nodes\}} and stores them in the vector \texttt{suspects}. Each suspect has a counter which represents the times a node is declared a suspect. Lastly, the controller checks if the counter of the suspect is equal to the number of its neighbors. In that case, the suspect is declared as attacker.   

\begin{algorithm} [t]
\caption{FDFF attackers identification}
\begin{algorithmic}
\State Wait alarms\{nodes\}
\State request graph\_information\{nodes\}
\For {n in alarms}
      \State {suspects = Extract\_neighbors(n)}
      \For {s in suspects}
          s\_counter++ 
          \If {s\_counter == total\_neighbors}
              \State s = attacker
          \EndIf
      \EndFor
\EndFor
\end{algorithmic}
\label{alg:1}
\end{algorithm}

\begin{figure*}[tb]
\centering
   \subfloat[Transmitting time\label{fig:detection_1_fdff_36_tx}]{%
      \includegraphics[width=0.49\textwidth]{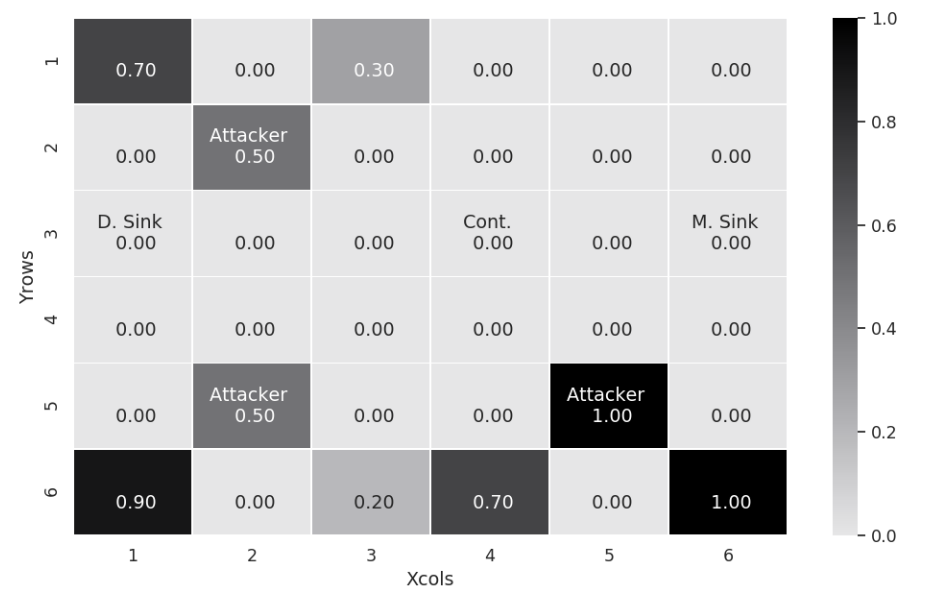}}
\hspace{\fill}
   \subfloat[Control packets received\label{fig:detection_1_fdff_36_rx_ctrl} ]{%
      \includegraphics[width=0.50\textwidth]{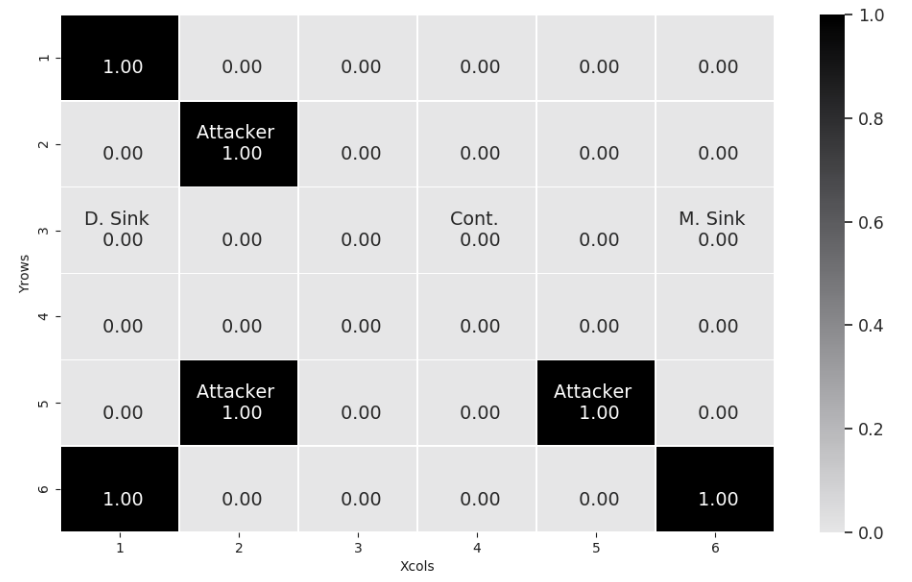}}
\caption{Attackers identification probability using Algoritmh \ref{alg:1} when the network is under an FDFF attack: case of 36 nodes. Each square in the map represents a node in the network. The number in the squares represent the probability of this node being classified as attacker.(a) shows the results when monitoring the transmitting time and (b) shows the results when monitoring the control packets received}
\label{fig:detection_1_fdff_36}
\end{figure*}

Fig. \ref{fig:detection_1_fdff_36} depicts the attacker identification results for 36 nodes when monitoring the transmitting time and the control packets received. The heat map shows the probability that each node has of being identified as attacker. We observed that for the case monitoring transmitting time (Fig \ref{fig:detection_1_fdff_36_tx}), in addition to the three attackers, seven benign nodes were identified as attackers as well. The probabilities of those nodes being misidentified as attackers ranged from $0.10$ to $1.00$, which means that some nodes were misidentified in all cases. In the case monitoring the control packets received (Fig. \ref{fig:detection_1_fdff_36_rx_ctrl}), all the attackers were correctly identified in all cases. On the other hand, 3 more nodes were misidentified in all the cases as well. 

We observed that the main problem of our identification algorithm was on the corners of the grid.\footnote{The reason for is because the corners have only two neighbors, and those neighbors are also in the attackers' neighborhood. This means, all the times our algorithm identified the attacker, automatically the corners were misidentified as attackers as well.} To solve this problem, we modified the suspects declaration in Algorithm \ref{alg:1}so that the node reporting also chooses one of its neighbors as suspect by inspecting the address of the node with the highest frequency of exchanges during the last ten samples. We chose ten samples because the slower detection when $\gamma=0$ is $1-S=0.84=9.6 \hspace{0.1cm} $samples in average (Fig. \ref{fig:1-s_FDFF_central}). Algorithm \ref{alg:2} shows the FDFF attacker identification algorithm after the modification.
\begin{algorithm} [t]
\caption{FDFF attackers identification 2}
\begin{algorithmic}
\State Wait alarms\{nodes\}
\State request graph\_information\{nodes\}
\For {n in alarms}
      \State {s = suspect(n)}
      \State  s\_counter++ 
          \If {s\_counter == total\_neighbors}
              \State s = attacker
          \EndIf
\EndFor
\end{algorithmic}
\label{alg:2}
\end{algorithm}
The results showed that the modification solved the misidenfitication problem. 

\begin{figure*}[htb]
\centering
   \subfloat[Transmitting time\label{fig:detection_2_fdff_36_tx}]{%
      \includegraphics[width=0.49\textwidth]{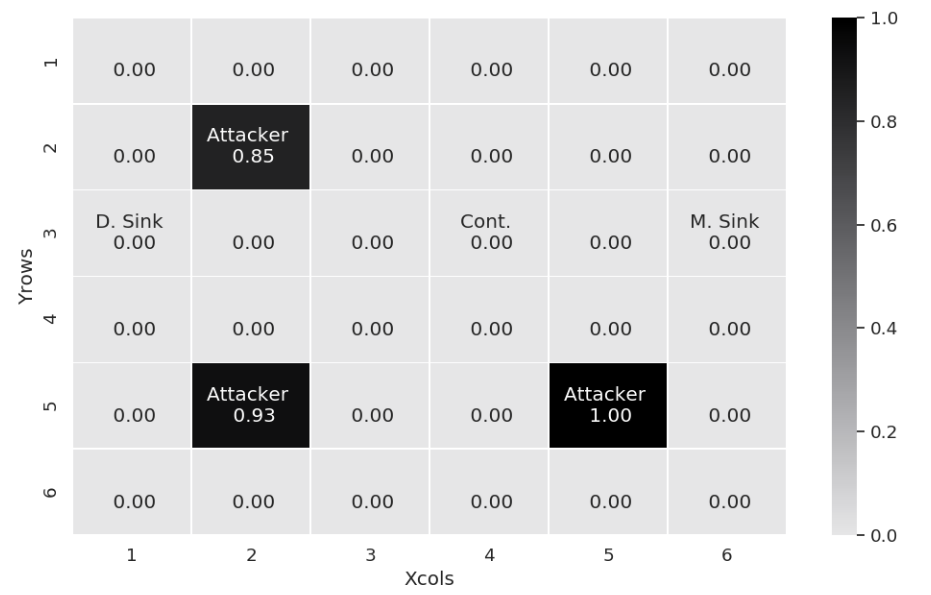}}
\hspace{\fill}
   \subfloat[Control packets received\label{fig:detection_2_fdff_36_rx_ctrl} ]{%
      \includegraphics[width=0.50\textwidth]{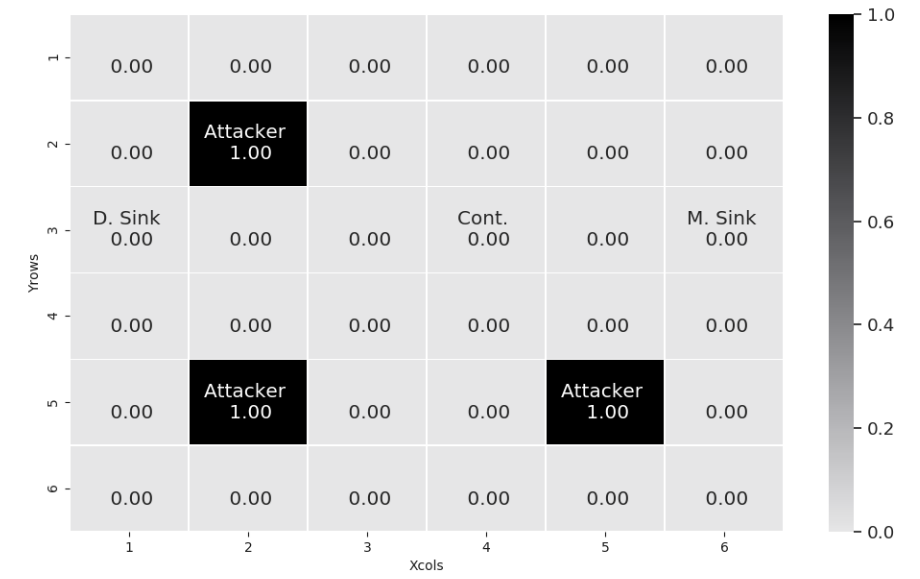}}
\caption{Attackers identification probability using Algorithm \ref{alg:2} when the network is under an FDFF attack: case of 36 nodes. Each square in the map represents a node in the network. The number in the squares represent the probability of this node being classified as attacker.(a) shows the results when monitoring the transmitting time and (b) shows the results when monitoring the control packets received}
\label{fig:detection_2_fdff_36}
\end{figure*}

\begin{figure*}[htb]
\centering
   \subfloat[Transmitting time\label{fig:detection_2_fdff_100_tx}]{%
      \includegraphics[width=0.49\textwidth]{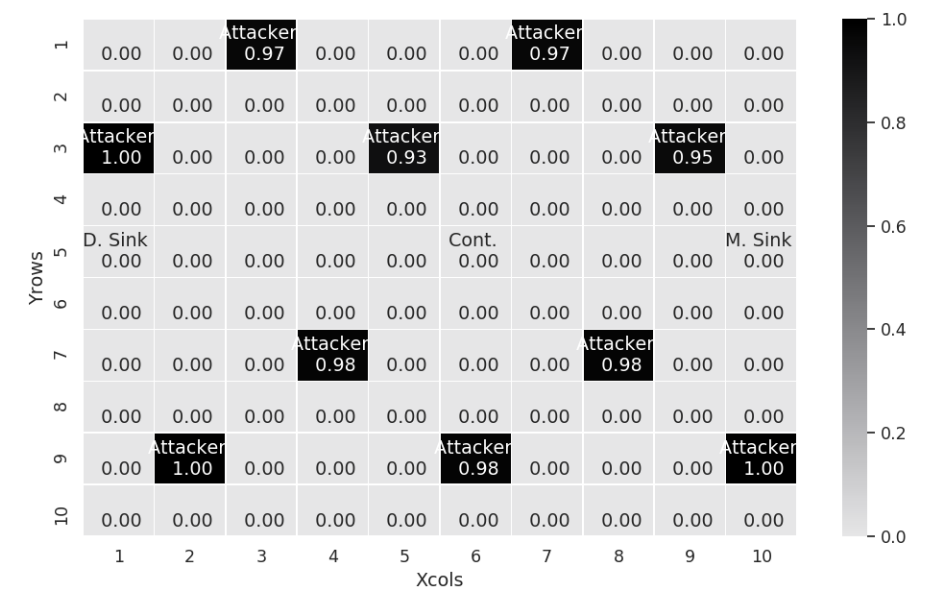}}
\hspace{\fill}
   \subfloat[Control packets received\label{fig:detection_2_fdff_100_rx_ctrl} ]{%
      \includegraphics[width=0.50\textwidth]{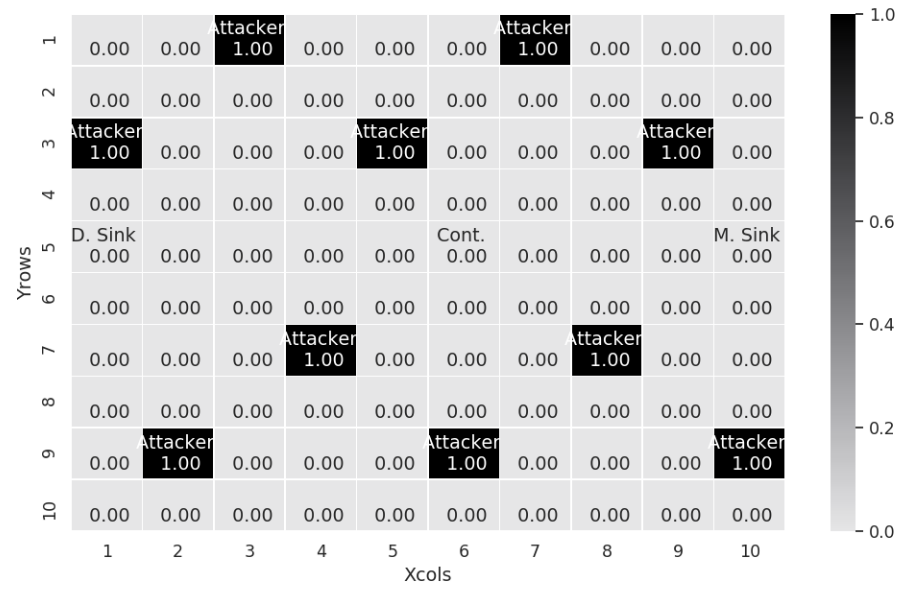}}
\caption{Attackers identification probability using Algorithm \ref{alg:2} when the network is under an FDFF attack: case of 100 nodes. Each square in the map represents a node in the network. The number in the squares represent the probability of this node being classified as attacker.(a) shows the results when monitoring the transmitting time and (b) shows the results when monitoring the control packets received}
\label{fig:detection_2_fdff_100}
\end{figure*}

In Fig. \ref{fig:detection_2_fdff_36} we observed that monitoring either the transmitting time or the control packets received, there were no misidentifications. When monitoring the control packets received the identification probability was $1.00$ for all the attackers, while when monitoring the transmitting time the identification probability was between $0.85$ and $1.00$. When evaluating the identification algorithm for 100 nodes (Fig. \ref{fig:detection_2_fdff_100}) we obtained excellent results as well; no misidentifications and identification probabilities over $0.93$. In fact, when monitoring the control packets received the identification probability was $1.00$ for all the attackers   

\begin{figure*}[t]
\centering
   \subfloat[36 nodes\label{fig:detection_1-s_fni_rx_ctrl_36}]{%
      \includegraphics[width=0.49\textwidth]{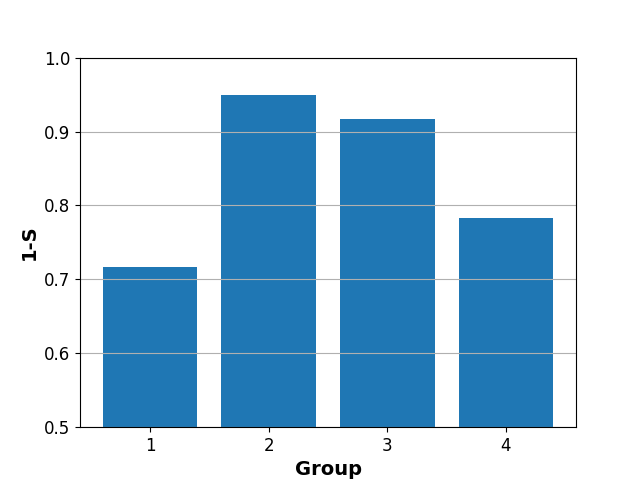}}
\hspace{\fill}
   \subfloat[100 nodes\label{fig:detection_1-s_fni_rx_ctrl_100} ]{%
      \includegraphics[width=0.50\textwidth]{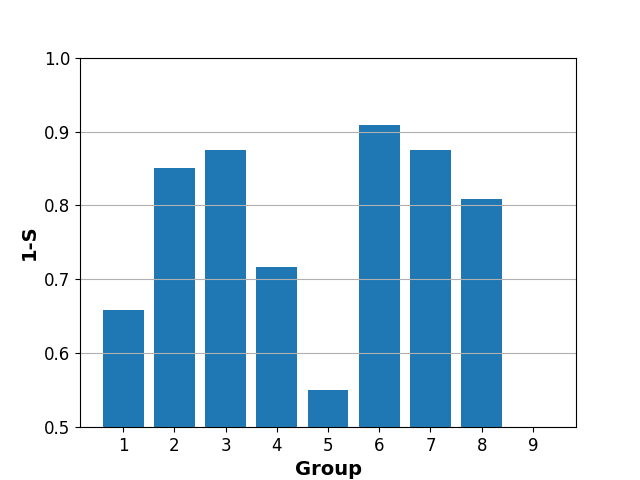}}
\caption{Detection speed (1-S metric) for FNI detection by data aggregation when monitoring the control packets received. (a) shows the results for 36 nodes and (b) shows the result for the case of 100 nodes}
\label{fig:detection_1-s_fni_rx_ctrl}
\end{figure*}

\subsection{Attacker detection in FNI attack} \label{sec:fni_attacker_detection}

The results in subsection \ref{sec:fni_prob} showed that for the case of FNI attacks, the percentage of nodes with high detection probability was low and also not all attackers' neighbors detected the attack, opposite to the observed for the FDFF attack. Because of this, we evaluated the attacker detection based on data aggregation. Our objective was to, at least, identify the area where the attacker was located. 

From Fig. \ref{fig:groups_fni_36_100} we noticed that our FNI detection strategy based on data aggregation increased the detection probability if compared with our initial approach, running the detector on every node. On the other hand, the data aggregation strategy results show detection probabilities over $37\%$ in areas without attackers. The first impression is that, in the case we track the attackers based on the alarm received from one group, this could lead to false positives because of the detection probability in areas without attackers. Thus, we analyzed the detection speed on every group. 

Fig. shows the $1-S$ metric (normalized $DMT$) for 36 and 100 nodes when monitoring the control packets received. For 36 nodes (Fig. \ref{fig:detection_1-s_fni_rx_ctrl_36}), our results showed that group 1 (the group without an attacker) has the lowest $1-S$, which means this is the last group reporting an alarm. %Now, suppose we trust on the hypothesis that the groups reporting an alarm has an attacker within it. We may execute a mitigation strategy targeting the firsts groups reporting an alarm because, as shown in Fig. \ref{fig:groups_fni_36_rx_ctrl}, these groups has the highest detection probability. In the best scenario, a successfully mitigation will stop the attack before the detectors on the groups without attackers detect a change.  
However, in the case of 100 nodes (Fig. \ref{fig:detection_1-s_fni_rx_ctrl_100}) the results did not show a similar trend. % on the first group reporting an alarm. Groups 4, 5, and 6 are the ones without an attacker within it; interestingly, group 6 had the highest $1-S$. Conversely, group 9 has one attacker and did not report alarms. This occurred because of two reasons: i) the attacker in group 9, which is in the border with group 6, had impact on group 6, and ii) the data sink in group 6 increased the packets traffic which accelerated the change in the time series. 

In conclusion, with respect to attacker identification, for the FDFF attach we proposed %Algorithm \ref{alg:1} to identify attackers when the network is under FDFF attack, but results showed this algorithm leads to false positive cases. Then, we proposed 
Algorithm \ref{alg:2} %and results showed it solved the false positive cases problem and 
that was shown to identify attackers with a probability over $0.93$ when monitoring the transmitting time, and a identification probability equal to $1$ when monitoring the control packets received. Conversely, for the FNI attack 
%For the FNI attack, we analyzed the attackers identification based on the data aggregation detection results when monitoring the control packets received. We noticed that for 36 nodes, the groups with attackers within it reported the alarm faster than the group without attackers. On the other hand, for 100 nodes 
we did not observe a reliable relation between any metric and the presence of attackers in the groups.

\section{Conclusion} \label{sec:conclusion}

In this work we proposed a centralized and a decentralized intrusion detection algorithm for WS-SDN constrained networks based on CP detection. The main strengths of our proposal is the high detection rates, the identification of the type of the attack and the localization or even identification of the attacker in some cases. The centralized approach provides a global view of the attack and allows us to identify the type of the attack; on the other hand the distributed detection provides information to identify the nodes launching the attack.

We evaluated our proposals through simulations using IT-SDN, Contiki-3.0 and the COOJA simulator, emulating Tmote sky motes. We simulated topologies of 36 and 100 nodes, varying the number of attackers in $5\%$, $10\%$, and $20\%$ of the total of nodes in the topology. We parameterized the centralized detector to either maximize the detection rate or the detection speed. %Then, we evaluated our proposal performance using the optimal parameters.
%From previous works we noticed that current proposals do not consider constrained networks resources, reported high detection rates only in small networks, and there is a lack of mechanisms to detect the type of the attack. 
%Our proposal was tested through simulation over a IEEE 802.15.4 network using resource constrained nodes. 
Our results showed detection rates over $96\%$ in networks of 36 and 100 nodes when using the centralized approach and were able to identify the type of the attack with a probability over $0.89$. Furthermore, we observed a FDFF attackers' identification with probability over $0.93$ when using the distributed detection.

As future work, we envisage to develop a full implementation of both approaches and compare their impact on the network performance and resource usage and to integrate both implementations to obtain the benefits of both approaches. Furthermore, we intend to explore the use of machine learning based fusion to tackle the identification of the attacker in the case of the FNI attack.

% or


% \appendices

% % you can choose not to have a title for an appendix
% % if you want by leaving the argument blank
% \section{}

\section*{Acknowledgment}
This study was financed in part by the Coordena\c{c}\~{a}o de Aperfei\c{c}oamento de Pessoal de N\'{i}vel Superior - Brasil (CAPES) - Finance Code 001 and by the ELIOT project (ANR-18-CE40-0030 / FAPESP 2018/12579-7).
Gustavo A. Nunez Segura is supported by Universidad de Costa Rica. 

\ifCLASSOPTIONcaptionsoff
  \newpage
\fi



%\bibliographystyle{IEEEtran}


% \end{thebibliography}
\bibliographystyle{IEEEtran}
\bibliography{IEEEfull,main}

\begin{thebibliography}{10}
\providecommand{\url}[1]{#1}
\csname url@samestyle\endcsname
\providecommand{\newblock}{\relax}
\providecommand{\bibinfo}[2]{#2}
\providecommand{\BIBentrySTDinterwordspacing}{\spaceskip=0pt\relax}
\providecommand{\BIBentryALTinterwordstretchfactor}{4}
\providecommand{\BIBentryALTinterwordspacing}{\spaceskip=\fontdimen2\font plus
\BIBentryALTinterwordstretchfactor\fontdimen3\font minus
  \fontdimen4\font\relax}
\providecommand{\BIBforeignlanguage}[2]{{%
\expandafter\ifx\csname l@#1\endcsname\relax
\typeout{** WARNING: IEEEtran.bst: No hyphenation pattern has been}%
\typeout{** loaded for the language `#1'. Using the pattern for}%
\typeout{** the default language instead.}%
\else
\language=\csname l@#1\endcsname
\fi
#2}}
\providecommand{\BIBdecl}{\relax}
\BIBdecl

\bibitem{Ieee2015}
D.~{Kreutz}, F.~M.~V. {Ramos}, P.~E. {Veríssimo}, C.~E. {Rothenberg},
  S.~{Azodolmolky}, and S.~{Uhlig}, ``{Software-Defined Networking: A
  Comprehensive Survey},'' \emph{Proc. IEEE Proc.}, vol. 103, no.~1, pp.
  14--76, Jan 2015.

\bibitem{Kobo2017}
H.~I. {Kobo}, A.~M. {Abu-Mahfouz}, and G.~P. {Hancke}, ``{A Survey on
  Software-Defined Wireless Sensor Networks: Challenges and Design
  Requirements},'' \emph{IEEE Access}, vol.~5, pp. 1872--1899, 2017.

\bibitem{8017556}
S.~{Bera}, S.~{Misra}, and A.~V. {Vasilakos}, ``Software-defined networking for
  internet of things: A survey,'' \emph{IEEE Internet of Things Journal},
  vol.~4, no.~6, pp. 1994--2008, 2017.

\bibitem{8104765}
S.~W. {Pritchard}, G.~P. {Hancke}, and A.~M. {Abu-Mahfouz}, ``Security in
  software-defined wireless sensor networks: Threats, challenges and potential
  solutions,'' in \emph{2017 IEEE 15th International Conference on Industrial
  Informatics (INDIN)}, 2017, pp. 168--173.

\bibitem{8377989}
F.~{Restuccia}, S.~{D’Oro}, and T.~{Melodia}, ``Securing the internet of
  things in the age of machine learning and software-defined networking,''
  \emph{IEEE Internet of Things Journal}, vol.~5, no.~6, pp. 4829--4842, 2018.

\bibitem{8215418}
S.~S. {Bhunia} and M.~{Gurusamy}, ``{Dynamic attack detection and mitigation in
  IoT using SDN},'' in \emph{27th Int. Telecommun. Netw. and Appl. Conf.
  (ITNAC)}, Nov 2017, pp. 1--6.

\bibitem{9090824}
Y.~{Jia}, F.~{Zhong}, A.~{Alrawais}, B.~{Gong}, and X.~{Cheng}, ``Flowguard: An
  intelligent edge defense mechanism against iot ddos attacks,'' \emph{IEEE
  Internet of Things Journal}, vol.~7, no.~10, pp. 9552--9562, 2020.

\bibitem{8993716}
N.~{Ravi} and S.~M. {Shalinie}, ``Learning-driven detection and mitigation of
  ddos attack in iot via sdn-cloud architecture,'' \emph{IEEE Internet of
  Things Journal}, vol.~7, no.~4, pp. 3559--3570, 2020.

\bibitem{8352645}
D.~{Yin}, L.~{Zhang}, and K.~{Yang}, ``{A DDoS Attack Detection and Mitigation
  With Software-Defined Internet of Things Framework},'' \emph{IEEE Access},
  vol.~6, pp. 24\,694--24\,705, 2018.

\bibitem{8998393}
C.~{Miranda}, G.~{Kaddoum}, E.~{Bou-Harb}, S.~{Garg}, and K.~{Kaur}, ``A
  collaborative security framework for software-defined wireless sensor
  networks,'' \emph{IEEE Transactions on Information Forensics and Security},
  pp. 1--1, 2020.

\bibitem{WANG2018119}
R.~Wang, Z.~Zhang, Z.~Zhang, and Z.~Jia, ``{ETMRM: An Energy-efficient Trust
  Management and Routing Mechanism for SDWSNs},'' \emph{Computer Networks},
  vol. 139, pp. 119 -- 135, 2018.

\bibitem{8805072}
R.~C.~A. {Alves}, D.~A.~G. {Oliveira}, G.~A. {Nunez Segura}, and C.~B. {Margi},
  ``{The Cost of Software-Defining Things: A Scalability Study of
  Software-Defined Sensor Networks},'' \emph{IEEE Access}, vol.~7, pp.
  115\,093--115\,108, Aug 2019.

\bibitem{OJIOT2019gnunez}
G.~A.~N. Segura, C.~B. Margi, and A.~Chorti, ``{Understanding the Performance
  of Software Defined Wireless Sensor Networks Under Denial of Service
  Attack},'' \emph{Open Journal of Internet Of Things (OJIOT)}, 2019, special
  Issue: Proc. Int. Workshop Very Large Internet of Things (VLIoT 2019) in
  conjunction with the VLDB 2019 Conf. Los Angeles, United States.

\bibitem{icc-2020}
N.~S. Gustavo, S.~Skaperas, A.~Chorti, L.~Mamatas, and B.~M. Cintia, ``{Denial
  of Service Attacks Detection in Software-Defined Wireless Sensor Networks},''
  in \emph{{SecSDN IEEE Int. Conf. Commun. (ICC)}}, Dublin, Ireland, Jun. 2020.

\bibitem{latincom-2020}
G.~A.~N. {Segura}, A.~{Chorti}, and C.~B. {Margi}, ``Multimetric online
  intrusion detection in software-defined wireless sensor networks,'' in
  \emph{2020 IEEE Latin-American Conference on Communications (LATINCOM)},
  2020, pp. 1--6.

\bibitem{Chorti-Hollanti}
A.~Chorti, C.~Hollanti, J.-C. Belfiore, and H.~V. Poor, ``Physical layer
  security: A paradigm shift in data confidentiality,'' in \emph{Physical and
  Data-Link Security Techniques for Future Communication Systems}, M.~Baldi and
  S.~Tomasin, Eds.\hskip 1em plus 0.5em minus 0.4em\relax Cham: Springer
  International Publishing, 2016, pp. 1--15.

\bibitem{McKeown2008}
N.~McKeown, T.~Anderson, H.~Balakrishnan, G.~Parulkar, L.~Peterson, J.~Rexford,
  S.~Shenker, and J.~Turner, ``{OpenFlow: Enabling Innovation in Campus
  Networks},'' \emph{SIGCOMM Comput. Commun. Rev.}, vol.~38, no.~2, pp. 69--74,
  Mar. 2008.

\bibitem{7226783}
I.~Ahmad, S.~Namal, M.~Ylianttila, and A.~Gurtov, ``{Security in Software
  Defined Networks: A Survey},'' \emph{IEEE Commun. Surveys Tuts.}, vol.~17,
  no.~4, pp. 2317--2346, Fourthquarter 2015.

\bibitem{SINGH2020509}
\BIBentryALTinterwordspacing
M.~P. Singh and A.~Bhandari, ``New-flow based ddos attacks in sdn: Taxonomy,
  rationales, and research challenges,'' \emph{Computer Communications}, vol.
  154, pp. 509 -- 527, 2020. [Online]. Available:
  \url{http://www.sciencedirect.com/science/article/pii/S0140366419313830}
\BIBentrySTDinterwordspacing

\bibitem{7593247}
D.~B. {Rawat} and S.~R. {Reddy}, ``Software defined networking architecture,
  security and energy efficiency: A survey,'' \emph{IEEE Commun. Surveys
  Tuts.}, vol.~19, no.~1, pp. 325--346, Firstquarter 2017.

\bibitem{ShinSeungwon}
\BIBentryALTinterwordspacing
S.~Shin and G.~Gu, ``Attacking software-defined networks: A first feasibility
  study,'' in \emph{Proceedings of the Second ACM SIGCOMM Workshop on Hot
  Topics in Software Defined Networking}, ser. HotSDN '13.\hskip 1em plus 0.5em
  minus 0.4em\relax New York, NY, USA: Association for Computing Machinery,
  2013, p. 165–166. [Online]. Available:
  \url{https://doi.org/10.1145/2491185.2491220}
\BIBentrySTDinterwordspacing

\bibitem{7534866}
S.~{Khan}, A.~{Gani}, A.~W. {Abdul Wahab}, M.~{Guizani}, and M.~K. {Khan},
  ``Topology discovery in software defined networks: Threats, taxonomy, and
  state-of-the-art,'' \emph{IEEE Communications Surveys Tutorials}, vol.~19,
  no.~1, pp. 303--324, 2017.

\bibitem{alexander-timeseries}
\BIBentryALTinterwordspacing
A.~Aue and L.~Horvath, ``Structural breaks in time series,'' \emph{Journal of
  Time Series Analysis}, vol.~34, no.~1, pp. 1--16, 2013. [Online]. Available:
  \url{https://onlinelibrary.wiley.com/doi/abs/10.1111/j.1467-9892.2012.00819.x}
\BIBentrySTDinterwordspacing

\bibitem{6380529}
A.~G. {Tartakovsky}, A.~S. {Polunchenko}, and G.~{Sokolov}, ``Efficient
  computer network anomaly detection by changepoint detection methods,''
  \emph{IEEE Journal of Selected Topics in Signal Processing}, vol.~7, no.~1,
  pp. 4--11, 2013.

\bibitem{1388279}
{Haining Wang}, {Danlu Zhang}, and K.~G. {Shin}, ``Change-point monitoring for
  the detection of dos attacks,'' \emph{IEEE Transactions on Dependable and
  Secure Computing}, vol.~1, no.~4, pp. 193--208, 2004.

\bibitem{8988163}
S.~{Skaperas}, L.~{Mamatas}, and A.~{Chorti}, ``Real-time algorithms for the
  detection of changes in the variance of video content popularity,''
  \emph{IEEE Access}, vol.~8, pp. 30\,445--30\,457, 2020.

\bibitem{8835019}
------, ``{Real-Time Video Content Popularity Detection Based on Mean Change
  Point Analysis},'' \emph{IEEE Access}, vol.~7, pp. 142\,246--142\,260, 2019.

\bibitem{FREMDT201474}
\BIBentryALTinterwordspacing
S.~Fremdt, ``Asymptotic distribution of the delay time in page's sequential
  procedure,'' \emph{Journal of Statistical Planning and Inference}, vol. 145,
  pp. 74 -- 91, 2014. [Online]. Available:
  \url{http://www.sciencedirect.com/science/article/pii/S0378375813002139}
\BIBentrySTDinterwordspacing

\bibitem{haleplidis2015software}
E.~Haleplidis, K.~Pentikousis, S.~Denazis, J.~H. Salim, D.~Meyer, and
  O.~Koufopavlou, ``{Software-defined networking (SDN): Layers and architecture
  terminology},'' Internet Research Task Force (IRTF), Tech. Rep., 2015.

\bibitem{poor2008quickest}
H.~V. Poor and O.~Hadjiliadis, \emph{Quickest detection}.\hskip 1em plus 0.5em
  minus 0.4em\relax Cambridge University Press, 2008.

\bibitem{Osterlind2006}
F.~{Osterlind}, A.~{Dunkels}, J.~{Eriksson}, N.~{Finne}, and T.~{Voigt},
  ``{Cross-Level Sensor Network Simulation with COOJA},'' in \emph{Proc. IEEE
  Conf. Local Comput. Netw. (LCN)}, Nov 2006, pp. 641--648.

\bibitem{Thamires}
T.~{Luz}, G.~{Nunez}, C.~{Margi}, and F.~{Verdi}, ``{In-network performance
  measurements for Software Defined Wireless Sensor Networks},'' in \emph{16th
  IEEE Int. Conf. Netw., Sens. and Control (ICNSC 2019)}, May 2019.

\bibitem{energest}
\BIBentryALTinterwordspacing
A.~Dunkels, F.~Osterlind, N.~Tsiftes, and Z.~He, ``Software-based on-line
  energy estimation for sensor nodes,'' in \emph{Proceedings of the 4th
  Workshop on Embedded Networked Sensors}, ser. EmNets '07.\hskip 1em plus
  0.5em minus 0.4em\relax New York, NY, USA: Association for Computing
  Machinery, 2007, p. 28–32. [Online]. Available:
  \url{https://doi.org/10.1145/1278972.1278979}
\BIBentrySTDinterwordspacing

\end{thebibliography}


\begin{IEEEbiographynophoto}{Gustavo A. Nunez Segura}
is a PhD candidate at Universidade de S\~ao Paulo. He received the M.Sc. degree (2018) in Electrical Engineering from Universidade de S\~ao Paulo and the B.Sc. in Electrical Engineering from Universidad de Costa Rica. His main research interests include energy consumption and security in wireless sensor networks and software-defined networking
\end{IEEEbiographynophoto}

\begin{IEEEbiographynophoto}{Arsenia Chorti}
is an Associate Professor in Communications and Networks at ETIS UMR8051, CY Univesrity, ENSEA, CNRS in France since 2017 and has served as a Lecturer at the University of Essex, UK from 2013 to 2017. She is a chartered engineer from the Technical Chambers of Greece since 2007, Senior IEEE member since 2020, a member of the IEEE P1951.1 Working Group on Smart Cities Standardization and of the IEEE INGR Working Group on Security. Between 2017-2020 she has served as a member of the IEEE Teaching Awards Committee. Her research interests include physical layer security and wireless communications, context awareness, root cause analysis and anomaly detection.

\end{IEEEbiographynophoto}

%\newpage

\begin{IEEEbiographynophoto}{Cintia Borges Margi}
obtained her Ph.D. in Computer Engineering at University of California Santa Cruz (2006) , and her Habilitation (Livre Docencia) (2015) in Computer Networks from the University of Sao Paulo. She is Associate Professor in the Computer and Digital Systems Engineering department at Escola Politecnica – Universidade de S\~ao Paulo (EPUSP) since 2015, where she started as Assistant Professor in 2010. During 2007-2010 she was Assistant Professor at Escola de Artes, Ciencias e Humanidades da Universidade de S\~ao Paulo (EACH-USP). Her research interests include: wireless sensor networks and software-defined networking.
\end{IEEEbiographynophoto}


%\vfill


\end{document}